\documentclass[aps,prb,twocolumn,tightenlines,floatfix]{revtex4-1}
\usepackage{natbib}
\usepackage{dcolumn}
\usepackage{bm}
\usepackage{makeidx}
\usepackage{enumerate}
\usepackage[pdftex]{graphicx}
\usepackage{amsmath}
\usepackage{amsfonts}
\usepackage{amssymb}
\usepackage{mathrsfs}
\usepackage{bbm}
\usepackage{subfigure}
\providecommand{\U}[1]{\protect\rule{.1in}{.1in}}
\begin{document}
\title{Anomalous proximity effects at the interface of $s$- and $s_{\pm}$- superconductors}
\author{Valentin G. Stanev and Alexei E. Koshelev}
\affiliation{Materials Science Division, Argonne National Laboratory, Argonne,
Illinois 60439}
\date{\today }

\begin{abstract}
We study proximity effects close to a boundary between $s$ and $s_{\pm}$
superconductors. Frustration, caused by interaction of the $s$-wave gap parameter
with the opposite-sign gaps of $s_{\pm}$ superconductor, leads to several anomalous
features. In the case of strong frustration a nontrivial time-reversal-symmetry
breaking (TRSB) state, with nonzero phase angles between all gap parameters, is
possible. In a more typical state, the $s$-wave
 order parameter is aligned with one of the $s_{\pm}$ gaps. The other
(anti-aligned) gap induces negative feature in the $s$-wave density of states, which
can serve as a fingerprint of $s_{\pm}$ state. Another consequence of the
frustration is an extended region in the parameter space in which $s$-wave
superconductivity is suppressed, despite being in contact with nominally stronger
superconductor. This negative proximity effect is always present for the TRSB state,
but extends even into the aligned states. We study these effects within a simple
microscopic model assuming dirty limit in all bands, which allows us to model the
system in terms of minimum number of the most relevant parameters. The described
anomalous features provide a route to establishing the possible $s_{\pm}$ state in
the iron-based superconductors.

\end{abstract}
\maketitle

\section{Introduction}

The discovery \cite{LaOFeAs} of the iron-based high-temperature superconductors has
brought new theoretical and experimental challenges to condensed matter physics.
Despite the undoubted progress in our knowledge (see, for example, Refs.
\onlinecite{physicaC, Paglione, Reviews}) some of the most intriguing questions
still do not have a definite answer. First among them by importance is the precise
form of the superconducting order parameter. The most plausible candidate so far is
the extended $s$-wave or $s_{\pm}$ state \cite{Mazin}. In this state the gaps on the
(hole-like) bands at the center and the (electron-like) bands at corners of the
Brillouin zone (BZ) have opposite signs. The physical origin of this state is in the
repulsive interband interactions, which likely dominate the pairing in these
materials. Although unconventional, such state from symmetry point of view is
indistinguishable from a conventional $s$-wave state, as they both belong to the
$A_{1g}$ representation of the lattice rotation group. Existence of such state is
supported by theoretical calculations done within the framework of several methods
-- Random Phase Approximation (RPA) \cite{Kuroki, Graser}, Functional
Renormalization Group (FRG) \cite{Fa Wang, Thomale}, Fluctuation Exchange (FLEX)
\cite{Rastko}, as well as analytic one-loop RG and diagrammatic calculations
\cite{Chubukov, Vlad, Maiti, Seo}.

In spite of a concentrated effort, the structure of the order parameter in these
superconductors has not yet been unambiguously established from experiment. The
angle resolved photoemission spectroscopy (ARPES) reveals uniform gaps on different
bands \cite{ARPES} for several compounds close to optimal doping, but it cannot
resolve the most crucial issue -- the relative sign of the gaps in the electron and
hole bands. This sign can be probed by some phase-sensitive experiments, similar to
ones performed for the cuprate high-temperature superconductors \cite{PhaseSens}.
Even though suggestions for such experiment have been made \cite{Phillips,Parker},
they have not been realized in practice yet \cite{PhaseSenseNote}.

The strongest support in favor of the $s_{\pm}$-state comes from inelastic neutron
scattering experiments \cite{MagRes}, which detect the emergence of a resonant
magnetic mode below the superconducting transition, as expected for such
sign-changing state. This mode was detected in almost all iron based superconductors
and its frequency scales approximately proportional to transition temperature
\cite{Paglione}. However, straightforward interpretation of the data is complicated
by the multiband character of the Fermi surface (FS) and the possible strong role of
interactions.

Another strong argument in favor of the $s_{\pm}$-state is microscopic coexistence
of antiferromagnetism and superconductivity experimentally demonstrated in some
compounds within a narrow doping range, most clearly in
Ba[Fe$_{1-x}$Co$_{x}$]$_{2}$As$_{2}$ \cite{LaplacePRB09,Fernandes10}. Spin-density
wave (SDW) has strong pair-breaking effect on the conventional $s_{++}$ state in
which the order parameter has the same sign in all bands. Such direct pair breaking
is absent if the order parameter has opposite signs in the bands connected by the
SDW ordering wave vector, as was confirmed by several theoretical studies
\cite{CoexTheory}. Thus, the SDW is much more compatible with the $s_{\pm}$ state
than with $s_{++}$ one.

Indirect probe of the bulk order parameter structure is provided by the
low-temperature behavior of the thermodynamic and transport properties which is
sensitive to the presence of quasiparticle states at the Fermi level. Extensive
studies have demonstrated a very rich behavior; all compounds fall into three
relatively well-defined groups. Several clean materials (LiFeAs \cite{LiFeAs},
Ba$_{1-x}$K$_{x}$Fe$_{2}$As$_{2}$\cite{BaKFeAS}) do not show low-energy
quasiparticles (they exhibit exponential temperature dependence of the London
penetration depth and no residual linear term in the specific heat), meaning that
all bands are fully gaped. Other materials, with weak impurity scattering, like
LaFePO \cite{LaFePO}, KFe$_{2}$As$_{2}$\cite{KFe2As2},
BaFe$_{2}$[As$_{1-x}$P$_{x}]_{2}$\cite{lambdaMatsuda}, shows behavior characteristic
of a clean superconductor with line nodes of the gap parameter, namely, linear
temperature dependence of the London penetration depth and square root magnetic
field dependence of the thermal conductivity. Moreover, for
BaFe$_{2}$[As$_{1-x}$P$_{x}$]$_{2}$ the presence of the node lines in one of the
bands has been directly confirmed by ARPES \cite{ARPESAsP}. Note that the existence
of accidental nodes does not necessarily contradict the overall picture of the
$s_{\pm}$ state -- even though the order parameter may change sign within one band,
what matters most is the sign of its average value over the Fermi surface of this
band. The third group of materials is formed by the compounds with strong scattering
by dopants, such as Ba[Fe$_{1-x}$Co$_{x}$]$_{2}$As$_{2}$, which have rather large
residual specific heat at low temperatures \cite{SpecHeat}, quadratic dependence of
the London penetration depth \cite{lambda}, and typically show residual thermal
conductivity \cite{ThCond}. These properties are not consistent with fully gapped
s-wave order parameter. While this behavior can be interpreted as indication for the
accidental nodes of the gap in combination with strong impurity scattering, it is
also compatible with $s_{\pm}$ state where interband impurity scattering generates
large number of subgap states and leads to finite density of states at the Fermi
level \cite{SubgapStates}.

The discovery of another iron-based superconducting family \cite{K122} -- the
chalcogenide 122's -- cast some doubts on the $s_{\pm}$ state as an universal state
for all compounds, since in these materials the hole bands around the $\Gamma$ point
are absent. Possible pairing mechanisms and structures of the order parameter in
view of recent experimental developments have been extensively discussed in recent
reviews \cite{Reviews}.
\begin{figure}[ht]
\includegraphics[width=0.48\textwidth]{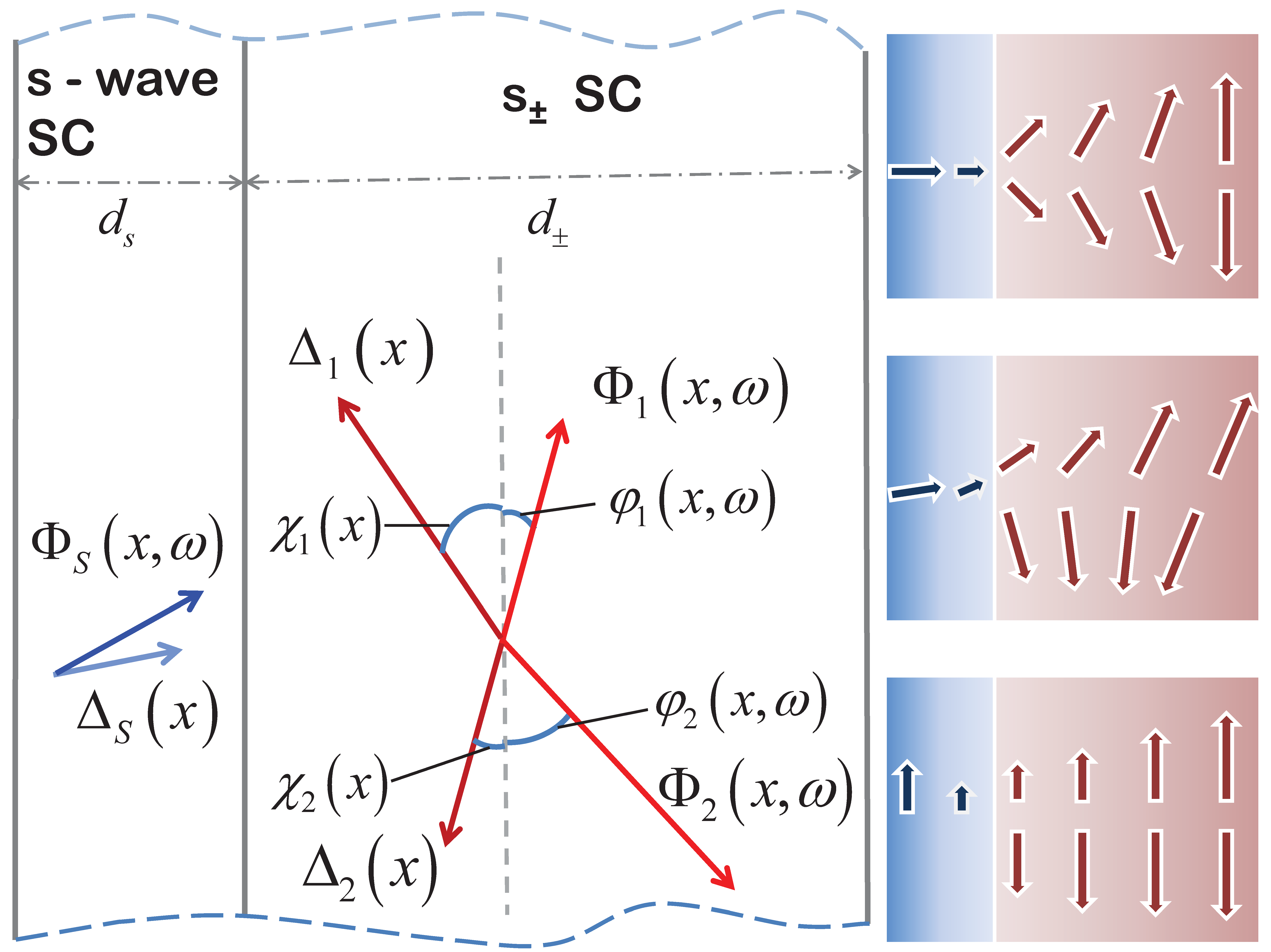}\caption{(Color online) \textit{On the left:}
The proximity structure composed of $s$-wave and $s_{\pm}$ superconductors
which we consider in this paper. The superconductors are described by the Green's
functions $\Phi$ depending on coordinate $x$ and Matsubara frequency $\omega$
and corresponding gap parameters $\Delta(x)$. For $s_{\pm}$ superconductor
these parameters depend on the band index $\alpha=1,2$, see Sec.\ \ref{Sec:Ueq} for details.
$\Phi_{s}$ couples to the $\Phi_{\alpha}$ on the $s_{\pm}$ side. In the case of
close coupling strengths with different bands, this creates
the possibility for a non-trivial relative phases $\phi_{\alpha}$ (as functions of $x$ and $\omega$), and
the TRSB state;
\textit{On the right:} The three possible states.
From top to bottom - the behavior of the anomalous part of the Greens'
functions in the symmetric and asymmetric TRSB state, and the aligned state.}
\label{SchemFig}
\end{figure}

In this paper we explore an alternative approach for probing the superconducting
order parameter in iron-based materials. It is based on proximity effect -- the
mutual influence of two superconductors, brought in contact. We study the vicinity
of a boundary between ordinary $s$-wave and $s_{\pm}$ superconductors, see Fig.\
\ref{SchemFig}. Competing interactions between the $s$-wave order parameter and
$s_{\pm}$ gaps with opposite signs lead to frustration. As a consequence of this
frustration, several interesting effects can appear, including the possibility of
new superconducting states. This is in contrast with proximity between conventional
(whether single- or multi-band) superconductors, which is a rather straightforward
phenomenon -- the phases of the gaps on both sides always align and their amplitudes
get closer (i.\ e., the smaller is enhanced and the larger is suppressed - see, for
example, Ref.\onlinecite{Golubov2}). In the case of a contact between $s$ and
$s_{\pm}$ superconductors there is no obvious way to align the phases and several
possibilities compete.

Various effects close to a boundary between $s$ and $s_{\pm}$ superconductors have
been already considered. Both phenomenological and microscopic methods were used,
number of interesting and novel results were obtained and different effects have
been suggested as possible fingerprints of the $s_{\pm}$ state. Several papers have
concentrated on the problem of a Josephson junctions between $s$ and $s_{\pm} $
superconductors\cite{Phillips,Parker,Linder,OtaPRL09} in the context of
SQUID\cite{Phillips} and Josephson\cite{Parker,Linder} interferometry, and
macroscopic quantum tunneling\cite{OtaPRL09, Ota} (for a detailed review of the
theoretical and experimental results see Ref.\ \onlinecite{Seidel}). In Refs.\
\onlinecite{Ng, VS, Berg,Lin} the mutual effects of $s$ and $s_{\pm}$
superconductors in contact were considered, and the possibility for a new
time-reversal symmetry breaking (TRSB) state close to the interface was discussed.
The methods used to obtain the TRSB state were different, however, all of them have
some intrinsic limitations and the obtained solutions cannot be regarded as fully
microscopic and self-consistent.

In this paper we consider in detail anomalous features of system composed of a
two-band $s_{\pm}$ superconductor in contact with a weaker (i.e., with smaller bulk
critical temperature) single-band $s$-wave superconductor, see Fig.\ \ref{SchemFig}.
We employ simple microscopic model, which assumes dirty limit in all
superconductors, but neglects interband scattering in $s_{\pm}$ superconductors.
This model allows us to describe the system using a minimum number of the most
essential and physically transparent parameters. Our approach neglects many features
of iron pnictides, such as the presence of more than two gaps, interband scattering,
orbital content of the bands, accidental nodes etc., which may be relevant for
detailed comparison of the theory with the experimental data. We believe, however,
that the overall picture we present will survive even in a more realistic
calculations.

As we show below, several non-trivial effects can be expected is such structures. If
the $s$-wave superconductor is much stronger coupled to one of the gaps on the
$s_{\pm}$ side than the other, it is natural to expect that its order parameter
aligns with this gap. The phase of the gap in the other band is then anti-aligned
with the $s$-wave gap and thus frustrated. This frustration leads to the possibility
of negative proximity effect -- the magnitudes of \emph{all} gaps being suppressed
close to the boundary (see the lower right panel on Fig.\ \ref{SchemFig}). This
effect is unique to the interface with $s_{\pm}$ order parameter, and has to be
contrasted with the case of $s$-$s_{++}$-wave structures, where the smallest gap is
always enhanced (we call this conventional proximity effect positive). As we
demonstrate below, in the aligned state both positive and negative effects are
possible.
We again emphasize that the negative proximity effect cannot be present for
conventional $s$-wave order parameter (assuming that the interface does not have
extrinsic pair-breaking surface effects). Thus, observation of such effect will be a
definitive prove of the presence of $s_{\pm}$ state in iron pnictides. The reverse
is not true -- positive proximity effect is possible for both $s_{\pm }$ and
$s_{++}$-wave superconductors.

When the properties of the two bands on the $s_{\pm}$ side are roughly the same, and
the coupling across the boundary is identical, even more interesting possibility
arises \cite{Ng, VS}. Since the frustration in such system is large, another state
emerges as a compromise. In it the relative phase of the gaps on the $s_{\pm}$ side
deviates from its bulk $\pi$ value close to the boundary. This tilting relieves some
of the frustration due to the inter-boundary coupling and disappears in the bulk
(see the upper right panel on Fig.\ \ref{SchemFig}). Such state is intrinsically
complex and thus breaks the time-reversal symmetry \cite{TRSBref}. Due to the finite
phase difference between the gaps, it is characterized by spontaneous supercurrents
with opposite direction flowing between the $s$-superconductor and $s_{\pm}$ bands
so that in ground state the total supercurrent is zero. In a more general situation
of different $s_{\pm}$ gaps, the TRSB state is possible when the Josephson coupling
energies between the $s$-wave order parameter and the opposite-sign bands exactly
compensate each other.
Another interesting consequence of our results is the possible phase transitions
between these different superconducting states. The direct way to  induce the TRSB
state is to vary the asymmetry. Unfortunately, there is no obvious way to do this.
More relevant experimentally are the transitions tuned by the temperature. Within
some range of parameters the aligned state is stable at higher temperature, but is
supplanted by the TRSB state at lower temperatures through a phase transition
\cite{KoshelevPhaseDiag}.

We focus on the situation when the direct interband reflection at the interface is
negligible, which is expected, for example, for the contact made at the [001]
surface of iron-based superconductor. In the opposite situation, when the interband
reflection is strong, a different kind of the TRSB state has been predicted
recently\cite{Bobkovi}.

In addition to these general effects we also consider the behavior of the density of
states (DoS) on both sides of the $s$-$s_{\pm}$ interface. This quantity can be
directly measured using the tunneling conductivity. As we show, in the aligned state
the aligned/anti-aligned gap induces positive/negative correction in the $s$-wave
DoS. Thus, observation of these features can be used to probe the multiband order
parameter (see also Ref.\ \onlinecite{ProxFingeprintEPL11}).

It is instructive to compare properties of the $s/s_{\pm}$ proximity system
considered in this paper with interfaces between $s$- and $d$-wave superconductors,
because $d$-wave represents a prominent case of sign-changing order parameter
realized in cuprate superconductors. The properties of the such interfaces have been
studied quite extensively, see, e.g., Refs.
\onlinecite{Buder90,Ohashi96,Lofwander04}, and several similar anomalous features
have been reported. In particular, for certain orientations the free surface of
$d$-wave superconductors or $s/d$ interface may generate $d+is$ or
$d_{x^2-y^2}+id_{xy}$ states with broken time-reversal symmetry
\cite{MatsumotoShiba,Lofwander04}. In contrast to the situation we consider here,
these states are induced by quasiparticle reflection at the interface between the
different-sign lobes. It was also found that the sign of proximity effect may be
negative for both superconductors \cite{Ohashi96}, but the conditions for this
phenomenon were not studied in detail.
The proximity corrections induced by the $d$-wave superconductor into the s-wave DoS
have been studied in Refs.\ \onlinecite{Ohashi96,Lofwander04}. The most prominent
feature is the peak near zero energy due to the Andreev bound state which splits
when the TRSB state is formed at the interface. Also, a smooth peak is typically
formed at the energy corresponding to the maximum gap of d-wave superconductor.

Let us outline the structure of the paper. First, in section \ref{FJJsn} we present
a very simple phenomenological model of a frustrated Josephson junction. In spite of
its simplicity, we believe that this model catches some of the essential physics of
the system. The results we obtain agree with the intuitive picture we presented
above: the TRSB state is stable when the Josephson couplings with the opposite-sign
bands are very close corresponding to strong frustration. Away from this region, the
aligned states are more favorable. The frustrated Josephson junction model predicts
continuous phase transitions between the aligned and TRSB states.

The above simple model, however, does not describe proximity effects, and misses
other important details as well. To develop a more realistic description of the
superconductivity on both sides of the interface, we use microscopic theory in the
dirty limit presented in section \ref{Sec:Ueq}, which describes the system by
multiband Usadel equations supplemented with the appropriate boundary conditions. In
section \ref{Sec:AnalytWeak} we present analytical results obtained in the limit of
weak coupling between the superconductors, and in section \ref{NumProc} we describe
the procedure used for numerical solution of the equations. Even though this
approach is strictly applicable only in the dirty limit, we expect our results to be
qualitatively (or even quantitatively) correct in the clean case as well.

Within the developed framework we study different superconducting states. We start
with the more conventional aligned state (in section \ref{AlSt}) and obtain both
positive and negative proximity effects, depending on the values of different
physical parameters. In section \ref{TRSB} we show that the TRSB state indeed
exists, and develop its quantitative description. Furthermore, the proximity effect
for such state is always negative (this is also clear from general considerations).
In section \ref{DaC} we summarize our results and discuss the possible limitations
of our approach.

\section{Simple model: frustrated Josephson junction}
\label{FJJsn}

To gain some insight into the phase diagram of the system let us first consider a
simple phenomenological model, which nevertheless catches essential physics --
Josephson junction between $s$ and $s_{\pm}$ superconductors (see also Ref.
\onlinecite{Lin}). The energy of such systems depends on the phase shifts between
the order parameters in the bands of the $s_{\pm}$ superconductor and the order
parameter in the $s$-wave superconductor, $\theta_{1,2}= \phi_{1,2}-\phi_{s}$. On
general grounds, the simplest form of this energy in reduced form can be written as
\begin{equation}
E(\theta_{1},\theta_{2})=\cos(\theta_{1}-\theta_{2})-t_{1}\cos\theta_{1}
-t_{2}\cos\theta_{2}
\end{equation}
with
\[
t_{\alpha}=E_{J,\alpha}/\mathcal{E}_{12}d_{\mathrm{\pm}}\ll1
\]
where $E_{J,\alpha}$ are Josephson coupling energies between the $s$-wave
superconductor and two bands of the $s_{\pm}$ superconductors, $\mathcal{E} _{12}$
is the interband coupling energy, and $d_{\mathrm{\pm}}$ is the thickness of
$s_{\pm}$ superconductor. For definiteness, we assume $t_{1}\leq t_{2}$. Minimizing
the energy, we get the equilibrium conditions:
\begin{subequations}
\begin{align}
-\sin(\theta_{1}-\theta_{2})+t_{1}\sin\theta_{1}  &  =0,\\
\sin(\theta_{1}-\theta_{2})+t_{2}\sin\theta_{2}  &  =0,
\end{align}
and, as a consequence,
\begin{equation}
t_{1}\sin\theta_{1}=-t_{2}\sin\theta_{2}.
\end{equation}
\end{subequations}
Excluding $\theta_{2}$, we obtain equation for $\theta_{1}$:
\[
\sin\theta_{1}\left(  -\sqrt{1-\frac{t_{1}^{2}}{t_{2}^{2}}\sin^{2}\theta_{1}}
-\cos\theta_{1}\frac{t_{1}}{t_{2}}+t_{1}\right)  =0.
\]
This equation has two solutions, corresponding to the aligned and TRSB states. For
the aligned state, $\sin\theta_{1}=0$ and we obtain $\theta_{1} =0,\ \theta_{2}=\pi$
and
\begin{equation}
E_{\mathrm{al}}=-1-t_{1}+t_{2}.\label{En-al}
\end{equation}
For the TRSB state we obtain
\begin{subequations}
\label{th-frustr}
\begin{align}
\cos\theta_{1}  &  =\frac{t_{2}}{2}+\frac{1}{2t_{2}}-\frac{t_{2}}{2t_{1}^{2}
},\\
\cos\theta_{2}  &  =\frac{t_{1}}{2}+\frac{1}{2t_{1}}-\frac{t_{1}}{2t_{2}^{2}}.
\end{align}
In particular, for the symmetric case $t_{1}=t_{2}$, and we have $\cos
\theta_{1}=\cos\theta_{2}=t_{1}/2$. The energy for the frustrated state can be
written as:
\end{subequations}
\begin{equation}
E_{\mathrm{fr}}=-\frac{1}{2}\left(  t_{1}t_{2}+t_{2}/t_{1}+t_{1}/t_{2}\right)
.\label{En-fr}
\end{equation}
The transition to the aligned state occurs when $\cos\theta_{1}=1$, giving
\begin{equation}
t_{2}=\frac{t_{1}}{1+t_{1}}\label{TransFrAl}
\end{equation}
and the aligned state is stable for $t_{2}<t_{1}/(1+t_{1})$. Converting to real
units, we conclude that for weak Josephson coupling the frustrated state is realized
in the region $|E_{J,1}-E_{J,2}|<E_{J}^{2}/(\mathcal{E}_{12}d_{\mathrm{\pm}})$.

\begin{figure}[ptb]
\includegraphics[width=0.45\textwidth]{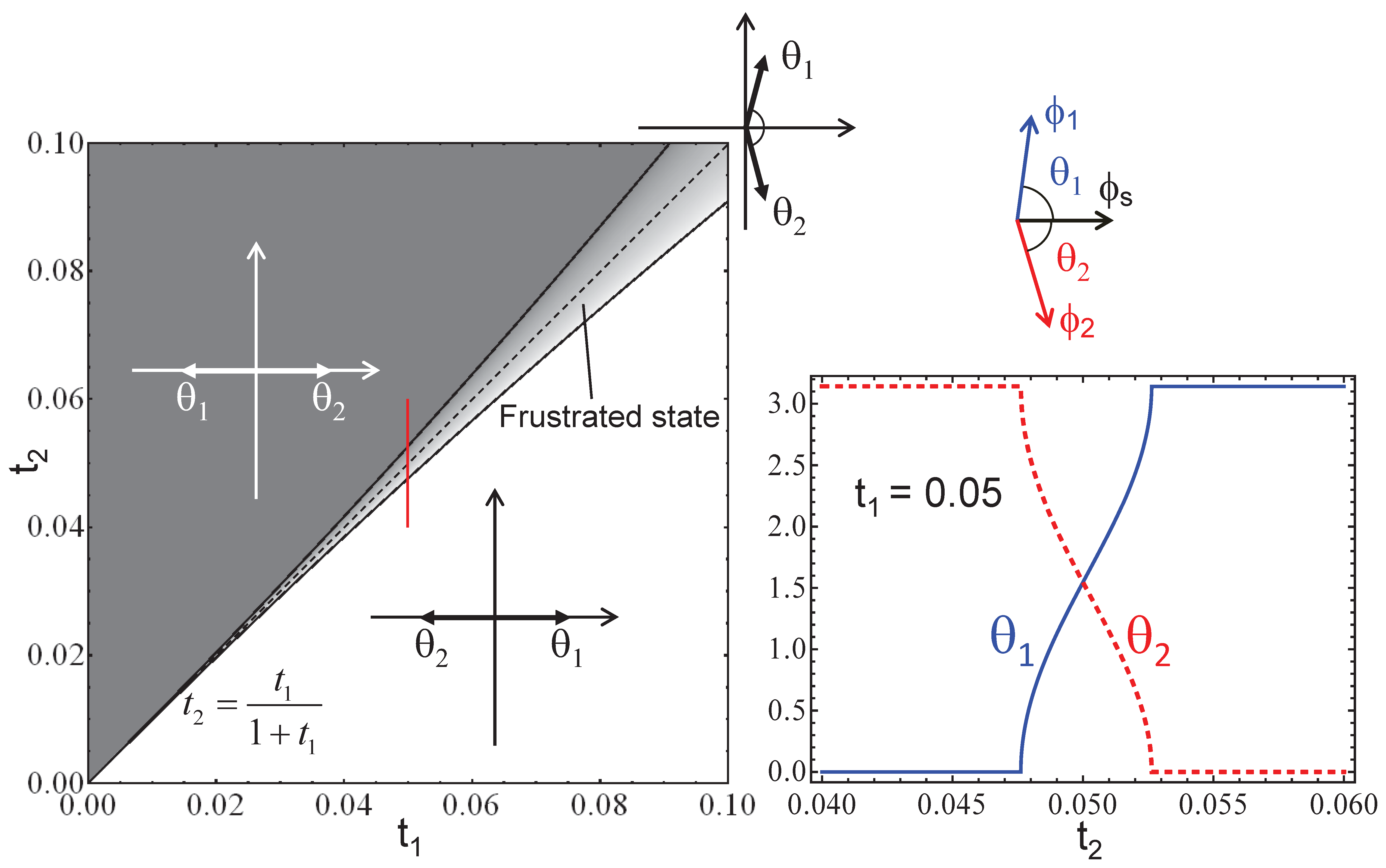}\caption{(Color online) \emph{Left
part:} The phase diagram of a frustrated Josephson junction. The TRSB state
exists near the diagonal part where $t_{1}\approx t_{2}$ . Away from this
diagonal region the aligned state is realized. \emph{ The right plot}
illustrates evolution of the phase angles $\theta_{1,2}$ along the vertical
line marked in the phase diagram.}
\label{FJJ}
\end{figure}

These results are summarized on Fig.\ \ref{FJJ}. As we have anticipated, the TRSB
state exists in the region where the frustration is maximum ($t_{1}\approx t_{2}$),
and is replaced by the aligned state when one of the couplings dominates. The
transition between the two states is continuous. Even though this simple model can
not pretend to quantitative description of the interface,  we expect that its
general features survive in a more microscopic setup, which we now proceed to
describe.

\section{Usadel equations and boundary conditions}
\label{Sec:Ueq}

We now move to a microscopic description of the problem. Let us consider a
``sandwich", consisting of a slab of a two-band $s_{\pm}$ superconductor with
thickness $d_{\mathrm{\pm}}$, in contact with a slab of a single-band $s$-wave
superconductor with thickness $d_{s}$, as shown on Fig.\ \ref{SchemFig}. We choose
the $x\!=\!0$ plane to be the boundary between the two layers. The main assumption
of our description is that both superconductors are in the dirty limit, but the
\emph{interband} scattering in the $s_{\pm}$ superconductor is negligible. In this
case superconductivity can be described by a simplified version of the Gor'kov
equations, known as Usadel equations \cite{Usadel}. The following formalism has been
developed in Ref. \onlinecite{Golubov1}, and applied to the case of conventional
two-band superconductors in Ref. \onlinecite{Golubov2}.

For the $s$-wave superconductor ($-d_{s}<x<0$), the equations for the impurity
averaged Greens' functions $G_{s}$ and $\Phi_{s}$ (where $\Phi_{s}=\omega
F_{s}/G_{s}$ and $F_{s}$ is the anomalous part of the single-particle Green's
function), with the necessary self-consistency equation, are:
\begin{subequations}
\begin{align}
&  \frac{D_{s}}{2\omega G_{s}}\left[  G_{s}^{2}\Phi_{s}^{\prime}\right]
^{\prime}-\Phi_{s} =-\Delta_{s},\\
&  2\pi T\ \sum_{\omega>0} \lambda_{s} \frac{G_{s}\Phi_{s}}{\omega}
=\Delta_{s},\ G_{s}=\frac{\omega}{\sqrt{\omega^{2}+|\Phi_{s}|^{2}}
}.\label{Ueqs}
\end{align}
where the prime denotes spatial derivative, and $\omega=2 \pi T (n+1/2)$ stands for
the Matsubara frequencies.

For the s$_{\pm}$-superconductor, $0<x<d_{\mathrm{\pm}}$ and the band index
$\alpha=1,2$, we have
\end{subequations}
\begin{subequations}
\begin{align}
\frac{D_{\alpha}}{2\omega G_{\alpha}}\left[  G_{\alpha}^{2}\Phi_{\alpha
}^{\prime}\right]  ^{\prime}-\Phi_{\alpha}  &  =-\Delta_{\alpha}
,\label{Ueqspm1}\\
2\pi T\sum_{\beta,\ \omega>0}\lambda_{\alpha\beta}\frac{G_{\beta}\Phi_{ \beta}
}{\omega}  &  =\Delta_{\alpha},\ G_{\alpha}=\frac{\omega}{\sqrt{\omega
^{2}+|\Phi_{\alpha}|^{2}}}.\label{Ueqspm2}
\end{align}
(we again emphasize the fact that interband impurity scattering has been neglected
-- for details see Ref. \onlinecite{Golubov2} and Section \ref{DaC}). We denote the
bulk critical temperatures of the $s_{\pm}$ and $s$-wave superconductors as $T_{c} $
and $T^{s}_{c}$, respectively. Since we consider $s_{\pm}$ superconductor, we assume
$\Delta_{1} \Delta_{2}<0 $ which is realized if $\lambda_{12},\lambda_{21}<0$. The
diffusion coefficients $D_{\{s, \alpha\}}$ are related to the conductivity
$\sigma_{\{s, \alpha\}}$ as $\sigma_{\{s, \alpha\}}=e^{2}\nu_{\{s, \alpha\}} D_{\{s,
\alpha\}}$, where $\nu_{\{s, \alpha\}}$ is the normal DoS. The ratio of the off-diagonal coupling constants is given by the
ratio of partial normal DoS, $\lambda_{\alpha \beta}/\lambda_{\beta
\alpha}=\nu_{\beta}/\nu_{\alpha}$. It is convenient to normalize all energy scales
($\omega$ and gaps on \emph{both} sides) to $\pi T_{c}$. We also introduce coherence
lengths $\xi_{\alpha}= \sqrt{D_{\alpha}/2 \pi T_{c}}$ and $\xi^{\ast}_{s}=\sqrt{
D_{s}/2 \pi T_{c}}$ (note that $\xi^{\ast}_{s}$ is related to the true bulk
coherence length of the $s$-wave superconductor by $\xi_{s}=\xi^{\ast}_{s}
\sqrt{T_{c}/T^{s}_{c}}$ ).

Since we consider an interface, these equations have to be supplemented with
appropriate boundary conditions. These connect the Green's functions and their
derivatives at the $x=0$ plane and can be written as \cite{KL}:
\end{subequations}
\begin{subequations}
\begin{align}
&  \xi^{\ast}_{s}G_{s}^{2}\Phi_{s}^{\prime} =\sum_{\alpha}\frac{\xi_{\alpha} }
{ \gamma_{\alpha}}G_{\alpha}^{2}\Phi_{\alpha}^{\prime},\label{BCgen1}\\
&  \xi_{\alpha} G_{\alpha}\Phi_{\alpha}^{\prime} = -\frac{1}{\gamma_{B \alpha}
} G_{s}(\Phi_{s}-\Phi_{\alpha}),\label{BCgen2}
\end{align}
for $\alpha=1,2$ with
\end{subequations}
\begin{equation}
\gamma_{\alpha}=\frac{\rho_{\alpha}\xi_{\alpha}}{\rho_{s}\xi^{\ast}_{s}},
\ \ \gamma_{B \alpha}=\frac{R_{B \alpha}} {\xi_{\alpha}\rho_{\alpha}},
\end{equation}
where $\rho_s$ and $\rho_{\alpha}$ are the bulk resistivities of the s-wave
superconductor and $\alpha$ band and $R_{B \alpha}$ is the boundary resistivity for
band $\alpha$. We can combine Eqs.\ \eqref{BCgen1}-\eqref{BCgen2} and get the useful
equivalent form of the boundary condition \ref{BCgen1},
\begin{align}
\xi^{\ast}_{s}G_{s} \Phi_{s}^{\prime} = \sum_{\alpha}\frac{1}{\tilde{\gamma}_{B\alpha}}
G_{\alpha}(\Phi_{\alpha}-\Phi_{s}).\label{BCgen3}
\end{align}
where we introduced the new interface parameters $\tilde{\gamma}_{B
\alpha}=\gamma_{B \alpha} \gamma_{\alpha}$ which we will use together with
$\gamma_{B \alpha}$. We also have to specify the conditions on the external
boundaries:
\begin{align}
\Phi_{s}^{\prime}(-d_{s})=0, \ \ \Phi_{\alpha}^{\prime}(d_{\mathrm{\pm}})=0.
\end{align}

Four parameters enter the boundary conditions and control the strength and the sign
of the proximity effect -- $\gamma_{1}$ and $\gamma_{2}$ depend on the bulk
properties of the materials, whereas $\gamma_{B 1}$ and $\gamma_{B 2}$ describe the
boundary itself \cite{Note-gammaBa}. The first two parameters determine the relative
strength of the proximity effect between the $s$-wave superconductor and the
$s_{\pm}$ bands. In particular, large $\gamma_{\alpha}$ implies that the $s$-wave
material is more metallic than the $\alpha$ band on the $s_{\pm}$ side, and thus
strongly influences it through proximity, while remaining weakly affected by this
band itself. For the ratio of these parameters we derive the following relation
\begin{equation}
\frac{\gamma_1}{\gamma_2}=\frac{\nu_2\xi_2}{\nu_1\xi_1}=
\frac{\lambda_{12}\xi_2}{\lambda_{21}\xi_1}.
\label{gamma-a-ratio}
\end{equation}
The parameter $\gamma_{B \alpha}$ is inversely proportional to the transparency of
the boundary for the $\alpha$ band. Estimating these parameters is not easy, but for
the case of iron-based materials (which are semi-metals) in contact with typical
conventional superconductor, we generally expect $\gamma_{\alpha}$ to be large.

To find the density of states, we have to perform analytical continuation of the
Green's functions to real energies $i\omega\rightarrow E +i\delta$. The normalized
DoS is related to the real-energy Green's function by the standard expression
\begin{align}
&  N_{\{s, \alpha\}}(E, x) =\mathrm{Re}[G_{\{s, \alpha\}}(E, x)]\nonumber\\
&  =\mathrm{Re}\left[  \frac{E}{\sqrt{E^{2}\!-\!\Phi_{\{s, \alpha\}}
(E,x)\Phi_{\{s, \alpha\}}^{\ast}(\!-\!E,x)}}\right] .\label{DoSGreens}
\end{align}
In the following section we present analytical results for the Green's functions and
gap parameters obtained in the limit of weak coupling between superconductors (large
$\gamma_{B\alpha}$) in the case of aligned state.

Establishing the model parameters which would describe real materials requires
experimental determination of electronic and scattering properties of the individual
bands. While this is a challenging task, in principle, this can be done using
ARPES\cite{ARPES,ARPESAsP}, quantum oscillations \citep{BandParam1}, or multiple-band fits of the magnetotransport \citep{BandParam2} and optical measurements \cite{BandParam3}.

\section{Analytical results for weak coupling}
\label{Sec:AnalytWeak}

In the case of the weak coupling between $s$ and $s_{\pm}$ superconductors,
$\gamma_{B\alpha}\! \gg\! 1$, the contact-induced corrections to the gaps and
Green's function can be treated as small perturbations, $\Delta_{\{s,\alpha \}}(x)=
\Delta_{\{s,\alpha\}0}+\tilde{\Delta}_{\{s,\alpha\}}(x)$ $\Phi_{\{s, \alpha\}}(x)=
\Delta_{\{s,\alpha\}0}+\tilde{\Phi}_{\{s, \alpha\}}(x)$. The small corrections
$\tilde{ \Phi}_{\{s, \alpha\}}(\omega,x)$ and $\tilde{\Delta} _{\{s, \alpha\}}(x)$
can be computed analytically in the linear order with respect to $1/\gamma_{B\alpha}
$. Similar calculation for several types of junctions using somewhat different
approach has been done in Ref.\ \onlinecite{GolubovKuprPZHETF05}. We consider here
only the case of aligned gaps. The computation details are presented in Appendix
\ref{App-WeakCoupl} and general results can be presented in the form of Fourier
expansions. For the $s$-wave superconductor
$\tilde{\Phi}_{s}(\omega,x)=\sum_{m=0}^{\infty}\tilde{ \Phi} _{s,m}(\omega)\cos
k_{m}x$ and $\tilde{\Delta}_{s}(x)=\sum_{m=0}^{\infty} \tilde{ \Delta}_{s,m}\cos
k_{m}x$ with $k_{m}=m\pi/d_{s}$ and the Fourier components are
\begin{subequations}
\begin{align}
\tilde{\Phi}_{s,m}(\omega) & =\frac{\tilde{\Delta}_{s,m}}{1+\xi_{s,\omega}
^{2}k_{m}^{2}}-\frac{(2-\delta_{m})\xi_{s,\omega}^{2}/(d_{s}\xi_{s}^{\ast} )
}{ 1+\xi_{s,\omega}^{2}k_{m}^{2}}\nonumber\\
& \times\sum_{\alpha}\frac{\sqrt{\omega^{2} +\Delta_{s0}^{2}}}{\sqrt{
\omega^{2}+\Delta_{\alpha0}^{2}}}\frac{\Delta_{s0} -\Delta_{\alpha0}}{\tilde{
\gamma}_{B\alpha}},\label{Phi-sm}\\
\tilde{\Delta}_{s,m} & =\!-\frac{2\pi T}{Z_{s,m}} \!\sum_{\alpha,\omega>0}
\frac{\omega^{2}}{\left(  \omega^{2}\!+\!\Delta_{s0}^{2}\right)  \sqrt{
\omega^{2}\!+\!\Delta_{\alpha0}^{2}} }\nonumber\\
&  \times\frac{\left(  2-\delta_{m}\right)  \xi_{s,\omega}^{2}/\left(  d_{s}
\xi_{s}^{\ast}\right)  }{1+\xi_{s,\omega}^{2}k_{m}^{2}}\frac{\Delta_{s0}
-\Delta_{\alpha0}}{\tilde{\gamma}_{B\alpha}},\label{Deltasm}\\
Z_{s,m} & = 2\pi T\!\sum_{\omega>0}\frac{1}{\left(  \omega^{2}\!+\!\Delta_{s0}
^{2}\right)  ^{3/2}} \left(  \Delta_{s0} ^{2}\!+\!\omega^{2}\frac
{\xi_{s,\omega}^{2}k_{m}^{2}}{1+\xi_{s,\omega} ^{2}k_{m}^{2}}\right) \nonumber
\end{align}
where $\xi_{s,\omega}^{2}= D_{s}/(2\sqrt{\omega^{2}+\Delta_{s0}^{2}})$.

For the $s_{\pm}$ superconductor the corresponding expansions are $\tilde{
\Phi}_{\alpha}=\sum_{m=0}^{\infty}\tilde{\Phi}_{\alpha,m}\cos q_{m} x$,$\
\tilde{\Delta}_{\alpha}=\sum_{m=0}^{\infty}\tilde{\Delta}_{\alpha,m}\cos q_{m}x$
with $q_{m}=m\pi/d_{\mathrm{\pm}}$. The Fourier components $\tilde{
\Phi}_{\alpha,m}$ and $\tilde{\Delta}_{\alpha,m}$ are given by somewhat cumbersome
but closed analytical formulas,
\end{subequations}
\begin{subequations}
\begin{align}
\tilde{\Phi}_{\alpha,m} &  =\frac{\tilde{\Delta}_{\alpha,m}}{1+\xi
_{\alpha,\omega}^{2}q_{m}^{2}}+\frac{\left(  2-\delta_{m}\right)  \xi
_{\alpha,\omega}^{2}/\left(  d_{\mathrm{\pm}}\xi_{\alpha}\right)  }
{1+\xi_{\alpha,\omega}^{2}q_{m}^{2}}\nonumber\\
&  \times\frac{\sqrt{\omega^{2}+\Delta_{\alpha0}^{2}}}{\sqrt{\omega^{2}
+\Delta_{s0}^{2}}}\frac{\Delta_{s0}-\Delta_{\alpha0}}{\gamma_{B\alpha}
},\label{Phi-alpham}\\
\tilde{\Delta}_{\alpha,m} &  =2\pi T\sum_{\beta,\omega>0}U_{m,\alpha\beta}
\frac{\omega^{2}}{\left(  \omega^{2}+\Delta_{\beta0}^{2}\right)  \sqrt{
\omega^{2}+\Delta_{s0}^{2}}}\nonumber\\
&  \times\frac{\left(  2-\delta_{m}\right)  \xi_{\beta,\omega}^{2}/\left(
d_{\mathrm{\pm}}\xi_{\beta}\right)  }{1+\xi_{\beta,\omega}^{2}q_{m}^{2}}
\frac{\Delta_{s0}-\Delta_{\beta0}}{\gamma_{B\beta}},
\end{align}
and $\xi_{\alpha,\omega}^{2}\equiv D_{\alpha}/(2\sqrt{\omega^{2}
+\Delta_{\alpha0}^{2}})$. Here the matrix $U_{m,\alpha\beta}$ is defined by the
following relations
\end{subequations}
\begin{align*}
\hat{U}_{m} &  =\frac{1}{D_{U}}
\begin{bmatrix}
w_{22}\!-\!\Sigma_{m,2} & -w_{12}\\
-w_{21} & w_{11}\!-\!\Sigma_{m,1}
\end{bmatrix}
,\\
D_{U} &  =-\Sigma_{m,2}w_{11}\!-\!\Sigma_{m,1}w_{22}+\Sigma_{m,1}\Sigma_{m,2}
\end{align*}
with
\[
\Sigma_{m,\alpha}\!=\!2\pi T\sum_{\omega>0}\left[  \frac{\omega^{2}}{ \left(
\omega^{2}\!+\!\Delta_{\alpha0}^{2}\right)  ^{3/2}}\frac{1}{ 1\!+\!\xi
_{\alpha,\omega}^{2}q_{m}^{2}}\!-\!\frac{1}{\omega}\right]  \!+\!\ln\frac
{1}{t}
\]
and $\hat{w}=\hat{\lambda}^{-1}-\lambda^{-1}\hat{I}$ being the degenerate matrix
whose components are given by relations
\begin{align*}
w_{11} &  =\frac{\sqrt{\lambda_{-}^{2}/4+\lambda_{12}\lambda_{21}}-\lambda
_{-}/2}{\det\lambda},\ w_{12}=-\frac{\lambda_{12}}{\det\lambda},\\
w_{22} &  =\frac{\sqrt{\lambda_{-}^{2}/4+\lambda_{12}\lambda_{21}}+\lambda
_{-}/2}{\det\lambda},
\end{align*}
where $\lambda_{-}\equiv\lambda_{11}-\lambda_{22}$ and $\det\lambda
\equiv\lambda_{11}\lambda_{22}-\lambda_{12}\lambda_{21}$.

\begin{figure}[ptb]
\includegraphics[width=0.48\textwidth]{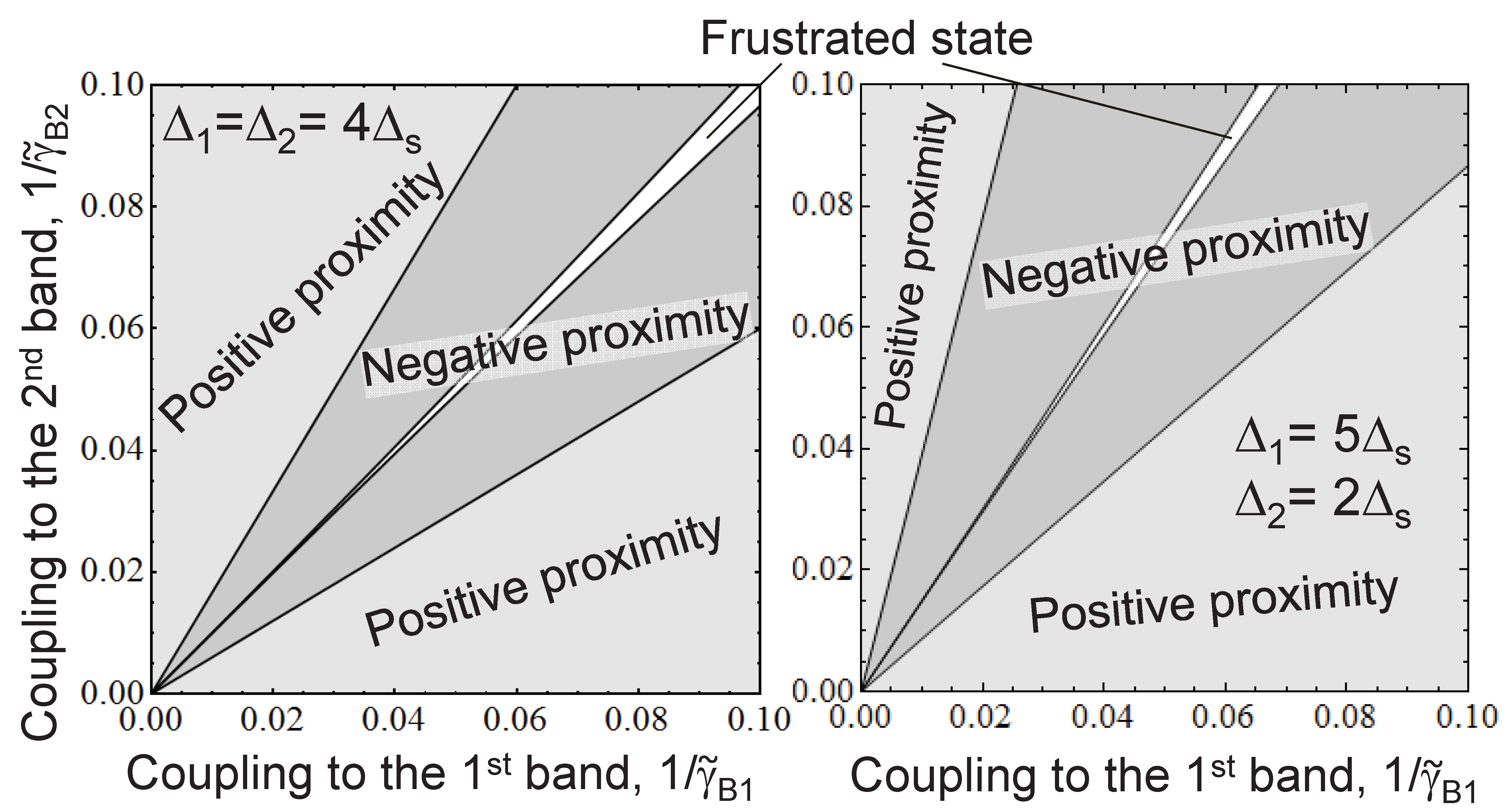}\caption{(Color online) The regions
of positive and negative proximity in the weak-coupling regime evaluated using
Eq.\ (\ref{AvDeltasLowT}). The left and right diagrams illustrate the cases of
identical and different $s_{\pm}$ gaps.}
\label{Fig-NegPosProx}
\end{figure}

The quantity $\tilde{\Delta}_{s,0}$ represents the average correction to the
$s$-wave gap parameter induced by the contact. In particular, the sign of
$\tilde{\Delta}_{s,0}$ determines wether $s$-wave superconductivity is enhanced or
suppressed (positive vs negative proximity effect). At low temperatures it is
possible to obtain an analytical result for the average gap correction, see Appendix
\ref{App-WeakCoupl},
\begin{align}
\frac{\tilde{\Delta}_{s,0}}{\pi T_{c}}  &  =\frac{\xi_{s}^{\ast}}{d_{s}}
\sum_{\alpha}U\left(\frac{\Delta_{s0}}{|\Delta_{\alpha0}|}\right)\frac{\Delta_{\alpha0}-
\Delta_{s0} }{\tilde{\gamma}_{B\alpha}|\Delta_{\alpha0}|}\label{AvDeltasLowT}
\\
\text{with }U(a)  &  =\frac{K(1\!-\!a^{2})-E(1\!-\!a^{2})}{1-a^{2}},\nonumber
\end{align}
where $K(m)=\int_{0}^{\pi/2}(1-m\sin^{2}\theta)^{-1/2}d\theta$ and
$E(m)=\int_{0}^{\pi/2}(1-m\sin^{2}\theta)^{1/2}d\theta$ are the complete elliptic
integrals. This general results further simplifies for the important particular case
$\Delta_{s0}\ll|\Delta_{\alpha0}|$. In this limit the elliptic integrals can be
expressed in terms of elemental functions leading to
\begin{equation}
\frac{\tilde{\Delta}_{s,0}}{\pi T_{c}}\approx\frac{\xi_{s}^{\ast}}{d_{s}}
\sum_{\alpha}\frac{\Delta_{\alpha0}-\Delta_{s0}}{\tilde{\gamma}_{B\alpha}
|\Delta_{\alpha0}|}\left[  \ln\left(  \frac{4|\Delta_{\alpha0}|}{\Delta_{s0} }
\right)  -1\right]  .\label{DsLimCase}
\end{equation}
From this result we can see that the partial contribution from the band is mostly
determined by the strength of coupling to this band $\propto1/\tilde{
\gamma}_{B\alpha}$. In addition, we observe that while the positive contribution
from the aligned band is proportional to the gap difference
$\Delta_{10}-\Delta_{s0}$, the negative contribution from the anti-aligned band is
proportional to the sum of the absolute gap values $|\Delta _{20}|+\Delta_{s0}$. As
a consequence, even in the aligned state the negative contribution may exceed the
positive one leading to the total negative proximity effect. This negative-proximity
region is especially broad in the case when the gap values $\Delta_{\alpha0}$ and
$\Delta_{s0}$ are close. Figure \ref{Fig-NegPosProx} illustrates regions of positive
and negative proximity in the coupling-constants plane,
$1/\tilde{\gamma}_{B1}$-$1/\tilde{\gamma}_{B2}$, obtained using Eq.\
(\ref{AvDeltasLowT}). The cases of identical and different $s_{\pm}$ gaps are
illustrated. We can see that in both cases the region of negative proximity occupies
significant region in the parameter space.

To find corrections to the densities of states, $\delta N_{s}(E)$ and $\delta
N_{\alpha}(E)$, we have to perform the analytical continuation $i\omega \rightarrow
E+i\delta$ in Eqs.\ (\ref{Phi-sm}) and (\ref{Phi-alpham}) and perform expansion in
Eq. (\ref{DoSGreens}) which gives
\begin{equation}
\delta N_{\{s, \alpha\}0}(E,x)=\mathrm{Re}\left[  \frac{E\Delta_{\{s,
\alpha\}0}^{\ast}\tilde{\Phi}_{\{s, \alpha\}}(E,x)}{\left(  E^{2}
-|\Delta_{\{s, \alpha\}0}|^{2}\right)  ^{3/2}}\right] .
\end{equation}
Simple analytical result illustrating general trends can be obtained at low
temperatures in the case of thin $s$-layer, $d_{s} < \xi_{s}$, and weak $s$-wave
superconductor, $|\Delta_{\alpha}|\gg|\Delta_{s}|$.\cite{ProxFingeprintEPL11} In
this case, the real-energy Green's function can be approximately evaluated as
\begin{equation}
\tilde{\Phi}_{s}(E,x)\!\approx\!\tilde{\Delta}_{s,0}-\pi T_{c}\frac{\xi
_{s}^{\ast}}{d_{s}}\sum_{\alpha}\frac{1}{\sqrt{\Delta_{\alpha0}^{2}\!- \!E^{2}
}}\frac{\Delta_{s0}\!-\!\Delta_{\alpha0}}{\tilde{\gamma}_{B\alpha}
}\label{PhisELim}
\end{equation}
Neglecting the first trivial term, we obtain the correction to the s-wave DoS
\begin{align}
\delta N_{s}(E,x)  &  \approx\pi T_{c}\frac{\xi_{s}^{\ast}}{d_{s}}\frac{
E\Delta_{s0} }{\left(  E^{2}-\Delta_{s0}^{2}\right)  ^{3/2}}\nonumber\\
&  \times\sum_{\alpha}\frac{\Delta_{\alpha0}-\Delta_{s0}}{\tilde{\gamma}
_{B\alpha}\sqrt{\Delta_{\alpha0}^{2}\!-\!E^{2}}}\Theta\left(  |\Delta_{\alpha0
}|\!-\!E\right).\label{sDoScorrWeak}
\end{align}
where $\Theta(x)$ is the step function.  We see that the aligned bands (positive
$\Delta_{\alpha}$) induce positive corrections to the $s$-wave DoS and the
anti-aligned bands (negative $\Delta_{\alpha}$) induces negative corrections. These
negative features can serve as definite fingerprint of $s_{\pm}$ state
\cite{ProxFingeprintEPL11}. Note that the perturbative result (\ref{sDoScorrWeak})
does not describes energy regions in the vicinity of the gap values,
$E\sim\Delta_{\alpha0}$.

To go beyond the weak-coupling regime we have to rely on numerical calculations. In
the following section we will describe numerical procedure and present results of
these calculations for different cases.

\section{Numerical procedure}
\label{NumProc}

For numerical modeling, it is convenient to use the so-called
$\theta$-parametrization, in which we write $G_{s}=\cos\theta_{s}$, $G_{\alpha}
=\cos\theta_{ \alpha}$. For $\Phi_{\alpha} $ and $\Phi_{s}$ two choices will prove
convenient. Let us first consider the (technically simpler) case of a significant
difference between the coupling of one of the two gaps on the $s_{\pm}$ side to the
gap on the $s$-wave side (weakly frustrated interface). As already mentioned in the
introduction, in such situation we expect the so-called aligned state to be stable.
In it the phase of $\Phi_{s}$ is aligned with one of $\Phi_{\alpha}$, while the
phase difference between $\Phi_{1}$ and $\Phi_{2}$ is $\pi$ -- this state obviously
belongs to the general class of $s_{\pm}$ states. Proximity effect on the $s$-wave
side enters through the (asymmetric) suppression of $\Delta_{1}$ and $\Delta_{2}$
close to the boundary (schematically shown on the lower right panel of Fig.\
\ref{SchemFig}).

In this state $\Phi_{\alpha}$ and $\Phi_{s}$ can be chosen real,
$\Phi_{\{s,\alpha\}}=\omega \tan\theta_{\{s,\alpha\}}$ (with the condition
$\Phi_{1}\Phi_{2}<0$). In the following, we assume that the coherence lengths of the
$s_{\pm}$ bands are equal $\xi_1=\xi_2$. We use dimensionless units: all energies
are normalized to $\pi T_{c}$ and the lengths on right/left are normalized to
$\xi^{\ast}_{s}/\xi_{\alpha}$. Separating the highest derivatives $\theta_{\{s,
\alpha\}} ^{\prime\prime}$, we can rewrite the Usadel equations as
\begin{equation}
\theta^{\prime\prime}_{\{s, \alpha\}} + \Delta_{\{s, \alpha\}} \cos
\theta_{\{s, \alpha\}} - \omega\sin\theta_{\{s, \alpha\}}=0,\label{UeqAlSt}
\end{equation}
This equation determines $\theta_{\{s, \alpha\}}$ as function of the discrete
Matsubara frequency $\omega=t(2n+1)$ with $t=T/T_c$.

For simplicity, we neglect the intraband pairing interactions and consider only the
repulsive interband coupling, parametrized by $\lambda_{12}$ and $\lambda_{21}$.
This is a reasonable approximation for the case of iron pnictides, in which the
interband pair-scattering is believed to be driving the superconductivity. Proper
generalization, including the (attractive or repulsive) intraband terms, is
straightforward. The self-consistency equation for $\Delta_{\alpha}$ \eqref{Ueqspm2}
becomes
\begin{equation}
\Delta_{\alpha}= - 2 t \lambda_{\alpha\beta} \sum_{n=0}^{N_{\mathrm{max}}}
\sin\theta_{\beta}.
\end{equation}
with $\alpha,\beta=1,2$ or $2,1$, $N_{\mathrm{max}}=\omega_{\mathrm{max}}/(2\pi
T)-1/2$, where $\omega_{ \mathrm{max}}$ is some frequency cutoff.

The boundary conditions at the outside boundaries, $x\!=\!-d_{s}, d_{\mathrm{\pm}}$,
are  $\theta^{\prime}_{s}(-d_{s})=0$ and
$\theta_{\alpha}^{\prime}(d_{\mathrm{\pm}})=0$. For the boundary conditions at the
$s$-$s_{\pm}$ interface we obtain in the case of the aligned state (see Appendix
\ref{App:BoundCondTheta})
\begin{subequations}
\begin{align}
&  \theta^{\prime}_{s}=\sum_{\alpha}\frac{\theta^{\prime}_{\alpha}}{
\gamma_{\alpha}},\label{BCal1}\\
&  \theta^{\prime}_{\alpha}=\frac{1}{\gamma_{B\alpha}}\sin(\theta_{\alpha}-
\theta_{s}).\label{BCal2}
\end{align}

Let us now consider the case of strongly frustrated boundary when the TRSB state
appears.  The Green's functions are now essentially complex quantities --
 $\Phi_{s} =\omega\tan\theta_{s} e^{i\varphi_{s}}$ and
$\Phi_{\alpha}= i \omega\tan\theta_{\alpha} e^{i \varphi_{\alpha}}$ (the factor $i$
in the definition of $\Phi_{\alpha}$ is for convenience). For the $s_{\pm}$ state we
have $\varphi_{1}-\varphi_{2}= \pi$. Correspondingly, the s-wave gap and the two
gaps of the order parameter can be written as $\Delta_s e^{i \chi_{s}}$ and $i
\Delta_{\alpha} e^{i \chi_{\alpha}}$, where the (real) parameters
$\Delta_{\{s,\alpha\}}$ and $\chi_{\{s,\alpha\}}$ have to be determined from the
self-consistency equations.

The transformed Usadel equations for the highest derivatives $\theta_{\{s,
\alpha\}}^{\prime\prime}$ and $\varphi_{\{s, \alpha\}}^{\prime\prime}$ in this case
become:
\end{subequations}
\begin{subequations}
\begin{align}
&\varphi ^{\prime\prime}  \!+  2 \cot\theta
\varphi^{\prime} \theta^{\prime}
 \!-\frac{\Delta}{\sin\theta}
 \sin(\varphi \!-\! \chi ) = 0 ,\label{Ue1}\\
&\theta^{\prime\prime} \! - \!   \sin\theta
\cos\theta (\varphi^{\prime})^{2} \!-\! \omega
\sin\theta \!+\!\Delta \cos\theta \cos(\varphi \!-\! \chi )\!=\!0,\label{Ue2}
\end{align}
where, for brevity, we omitted the subscripts $\{s, \alpha\}$. We use the same
normalization for all energies and lengths as in Eq.\ \eqref{UeqAlSt}.

The self-consistency equation for $\Delta_{\alpha}$ is:
\end{subequations}
\begin{align}
\Delta_{\alpha} e^{i \chi_{\alpha}}= - 2 t \lambda_{\alpha\beta} \sum
_{n=0}^{N_{\mathrm{max}}} \sin\theta_{\beta} e^{i \varphi_{\beta}}.
\end{align}
This complex equation can be split into two real equations for $\Delta _{\alpha}$
and $\chi_{\alpha}$:
\begin{align}
&  \Delta_{\alpha} =\nonumber\\
&  2 t \lambda_{\alpha\beta} \left[  \left(  \sum_{n}\sin\theta_{\beta}
\cos\varphi_{\beta}\right)  ^{2} \!+\! \left(  \sum_{n}\sin\theta_{\beta}
\sin\varphi_{\beta}\right)  ^{2} \right]  ^{1/2},\nonumber\\
&  \chi_{\alpha} = -\arctan\left(  \sum_{n}\sin\theta_{\beta} \sin
\varphi_{\beta}/\sum_{n}\sin\theta_{\beta} \cos\varphi_{\beta}\right).
\nonumber
\end{align}
The negative sign in the second equation -- a straightforward consequence of the
fact that $\lambda_{\alpha \beta}$ is repulsive -- leads to somewhat
counterintuitive result. The proximity effect tends to align the phases of
$\Phi_{s}$ and $\Phi_{\alpha}$. $\Phi_{\alpha}$, in turn, determines the order
parameter $\Delta_{\beta}$ through the self-consistency equation, which tends to
make the phases difference between $\Phi_{\alpha}$ and $\Delta_{\beta}$ close to
$\pi$ (to compensate for the ``wrong'' sign of the interaction). This means that,
due to the proximity effect, the phase difference between $\Delta_{\alpha}$ and
$\Delta_{s}$ \emph{increases} (measured from the positive axis), rather than
decreases -- see Fig.\ \ref{SchemFig} for a pictorial presentation of this argument.
This is a somewhat oversimplified picture, since $\varphi_{\alpha}$ depends on the Matsubara
frequency. As can be seen from the Eqs. \eqref{Ueqspm1}-\eqref{Ueqspm2}, for high
frequencies the derivative term becomes unimportant and $\Phi_{\alpha}
\rightarrow\Delta_{\alpha}$ meaning that $\varphi_{\alpha}
\rightarrow\chi_{\alpha}$. This implies that $\varphi _{\alpha}$ changes its sign as
a function of $\omega$. To obtain $\Delta_{\beta}$ we have to sum over all $n$ and
this leads to ``smearing'' of the contributions to $\chi_{\beta}$ over a phase
interval. Nonetheless, as we will see, the intuitive argument above seems to be
qualitatively correct.

The boundary conditions at the outside boundaries are given by
\begin{align*}
&  \theta^{\prime}_{s}(-d_{s})=0,\ \ \varphi^{\prime}_{s}(-d_{s})=0,\\
&  \theta_{\alpha}^{\prime}(d_{\mathrm{\pm}})=0,\ \ \varphi_{\alpha}^{\prime}(d_{\pm
})=0.
\end{align*}
The boundary conditions at the $s$-$s_{\pm}$ interface in $\theta$-parametrization
are more complicated and we derive them in Appendix \ref{App:BoundCondTheta}. In
reduced units these conditions are
\begin{subequations}
\begin{align}
&\theta _{\alpha }^{\prime }\!=\!-\frac{1}{\gamma _{B\alpha }}
\left[ \cos \theta _{\alpha }\sin \theta _{s}\sin \left( \varphi
_{s}\!-\!\varphi _{\alpha }\right)\! -\!\cos {\theta _{s}}\sin \theta _{\alpha }
\right],  \\
&\varphi _{\alpha }^{\prime }\sin \theta _{\alpha } =\frac{1}{\gamma _{B\alpha }}
\sin \theta _{s}\cos \left( \varphi _{s}-\varphi _{\alpha
}\right),\\
&\theta _{s}^{\prime }\! =\!-\!\sum_{\alpha }\frac{1}{\tilde{
\gamma}_{B\alpha }}\left[ \cos {\theta _{\alpha }}\sin \theta _{s}\!+\!\cos
\theta _{s}\sin \theta _{\alpha }\sin \left( \varphi _{\alpha }\!-\!\varphi
_{s}\right) \right],  \\
&\varphi _{s}^{\prime }\sin \theta _{s} =\sum_{\alpha }
\frac{1}{\tilde{\gamma}_{B\alpha }}\sin \theta _{\alpha }\cos \left( \varphi
_{\alpha }-\varphi _{s}\right).
\end{align}
\label{BoundCondThetaComplex}
\end{subequations}

To summarize, we have developed an entirely self-consistent scheme, based on Usadel
equations and supplemented by the appropriate self-consistency equations and
boundary conditions. Unfortunately, due to its considerable complexity, in the
general case it has to be solved numerically. To do this, we start with a guess for
$\theta_{s}$, $\theta_{\alpha }$ and $\varphi_{\alpha }$, solve it on the $s_{\pm}$
side, using two of the boundary conditions at $x=0$ and the boundary conditions at
$x=d_{\mathrm{ \pm}}$, then write the third condition at $x=0$ with the new
solutions for $\theta_{\alpha }$, $\varphi_{\alpha }$ and use it to solve for
$\theta_{s}$. With such obtained solution we re-write the two initially used
boundary conditions and again solve for $\theta_{\alpha}$ and $\varphi_{\alpha}$. We
repeat the process until self-consistency is achieved. To simplify the calculations,
in some cases we expanded the equations around the bulk value $\theta_{\alpha}=
\arctan(\Delta_{\alpha,0}/\omega)$, $\varphi_{\alpha}=0$, where $\Delta_{\alpha,0}$
is the bulk gap for $\alpha$ band.

Once calculations have produced self-consistent solutions of the equations on both
sides of the boundary, we can obtain the DoS, $N_{\{s, \alpha \}}(E,x)$. For this,
we rewrite the equations and boundary conditions with analytically continued
frequency $i\omega\rightarrow E +i\delta$, and with computed gap functions
$\Delta_{s}(x)$, $\Delta_{\alpha}(x)$ and solve them again for the real-energy
Green's functions $\theta_{\{s, \alpha \}}(E, x)$. Once these equations are solved,
the DoSs given by Eq.\ \eqref{DoSGreens} or, in the $\theta$-parametrization, by
\[
N_{\{s, \alpha\}}(E, x)\!=\!\mathrm{Re}[\cos{\theta_{\{s, \alpha\}}(E,
x)}].\nonumber
\]
In the following sections we well review properties of the proximity system for
different cases.

\section{Aligned state}
\label{AlSt}

With the formalism described in the sections \ref{Sec:Ueq}, \ref{Sec:AnalytWeak} and
\ref{NumProc}, we are now ready to study the proximity effects in several particular
cases.

\subsection{Identical $s_{\pm}$ bands}

The aligned states appear when the bulk and boundary condition parameters are not
symmetric with respect to the band indices interchange (asymmetric $s_{\pm}$ state).
For simplicity, we first consider the asymmetry only in the coupling strength
$\gamma_{B1}\! \neq\!\gamma_{B2}$, while keeping the rest of the parameters
symmetric. In particular, this means that close to the boundary $|\Delta_{1}|\neq|
\Delta_{2}|$, but in the bulk the symmetric $s_{\pm}$ state is restored and
$\Delta_{2}=-\Delta_{1}$ (see right middle panel of Fig.\ \ref{SchemFig}). In all
numerical calculations presented in this section, unless stated otherwise, we fixed
several parameters: $d_{\mathrm{\pm}}=8 \xi_{\alpha}$, $T^{s}_{c}=0.222 T_{c}$,
$T=0.1 T_{c}$. The choice of $T^{s}_{c}$ gives relation for the gap $\Delta
_{s}^{\mathrm{bulk}}/(\pi T_{c})=0.1213$. Other parameters which enter the
calculations are specified in the figure captions.

Since in a way the aligned state is closer to a conventional multiband
superconductor than the TRSB state is, we expect the usual positive proximity effect
to be predominant. Indeed, as we will see, the $s$-wave superconductor's gap tends
to be \emph{enhanced} by the presence of the stronger $s_{\pm}$ superconductor
(which in turn is suppressed). This is easy to understand when the $s$-wave gap is
much stronger coupled to one of the gaps on the $s_{\pm}$ side, since this case can
be thought as proximity between two single-band superconductors. Coupling to the
other gap on the $s_{\pm}$ side can be treated as a small perturbation. However,
when the interboundary couplings are close, and the system is strongly frustrated
(and thus close to the TRSB state), the proximity effect turns negative, with all
superconducting gaps suppressed close to the $x\!=\!0$ plane. Therefore the ``sign''
of the proximity effect on the $s$-wave side in the aligned state is not universal,
and depends on both bulk and boundary properties of the materials.
\begin{figure}[ptb]
\includegraphics[width=0.5\textwidth]{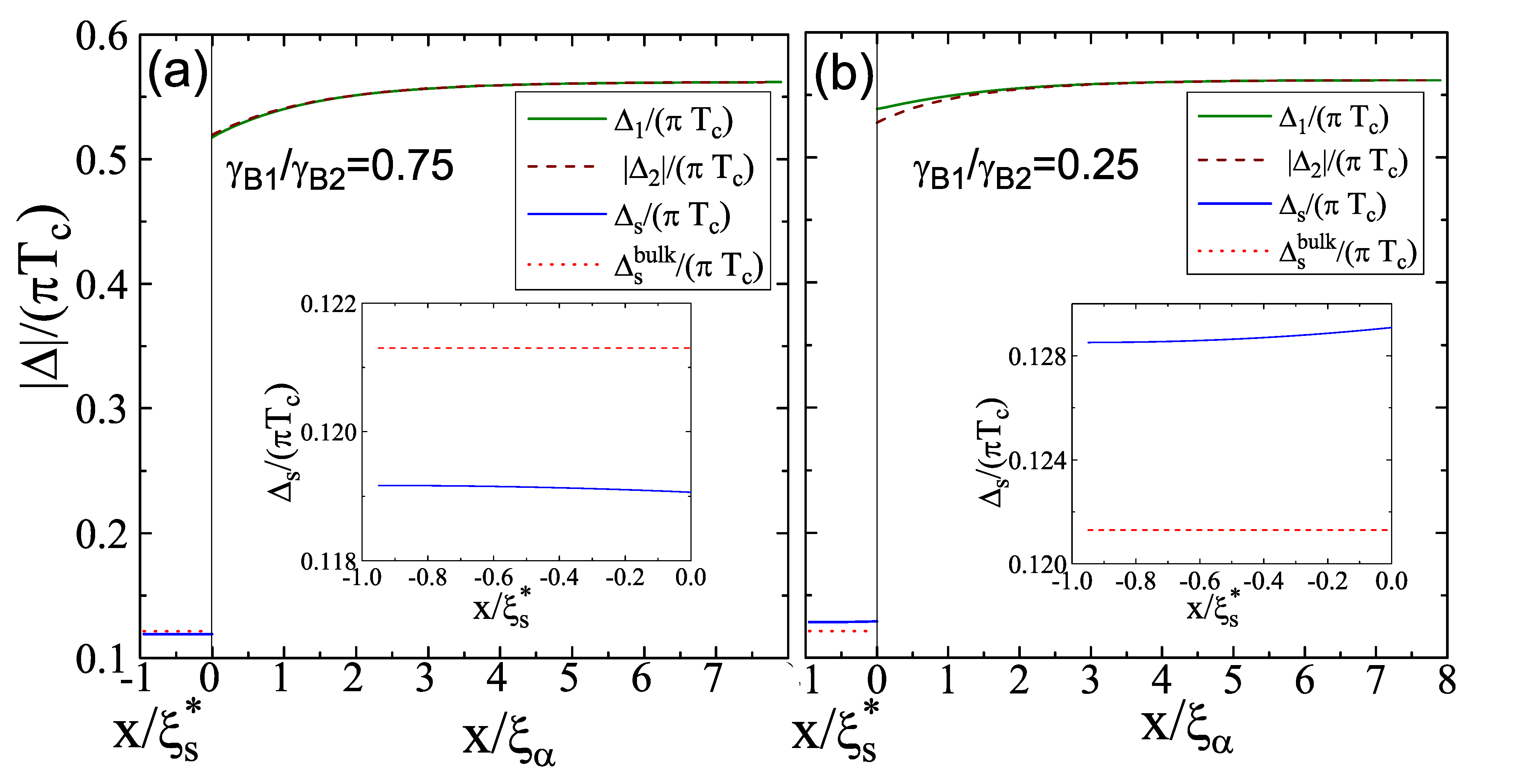}
\caption{(Color online) These plots illustrate behavior of the gaps in the proximity sandwich
for two values of the ratio $\gamma_{B1}/\gamma_{B2}$. The insets show blowups
of $\Delta_s(x)$ and, for comparison, $\Delta^{\mathrm{bulk}}_{s}$ is also
shown. Other used parameters are $\gamma_{B1}=10$, $d_{s}=\xi^{\ast}_{s}$, and
$\gamma_{\alpha}=10$. One can see that for weakly asymetric coupling
$\gamma_{B1}/\gamma_{B2}=0.75$ in the plot (a) the proximity effect is negative ($\Delta_s(x)<
\Delta^{\mathrm{bulk}}_{s}$) while for strongly asymetric coupling
$\gamma_{B1}/\gamma_{B2}=0.25$ in the plot (b) the proximity effect becomes positive
($\Delta_s(x)>\Delta^{\mathrm{bulk}}_{s}$). } \label{DSaligned1}
\end{figure}

On Fig.\ \ref{DSaligned1} we illustrate this effect by showing calculations of
$\Delta_{s}(x)$ for two different values of $\gamma_{B1}/\gamma_{B2}$ (we change
$\gamma_{B2}$ while keeping all other parameters of the system fixed). As can be
seen, $\Delta_{s}$ is enhanced or suppressed close the interface (corresponding to a
sign change of $\theta^{\prime}_{s}(0)$), depending on the ratio of the boundary
transparencies. Note that the transition between the two cases coincides with change
of the ratio of $\Delta_{1}(0)/|\Delta_{2}(0)|$. For positive (negative) proximity
effect this ratio is larger (smaller) than $1$. This is easy to understand from the
first condition in Eqs. \eqref{BCal1} -- it is clear that (for
$\gamma_{1}\!=\!\gamma_{2}$) the sign of $\theta^{\prime}_{s}$ is determined by the
$\theta^{\prime}_{1}+\theta^{\prime}_{2}$. Also remember that $\Delta_{1}$ is
determined by $\theta_{2}$ and vice versa.
\begin{figure}[ptb]
\includegraphics[width=0.5\textwidth]{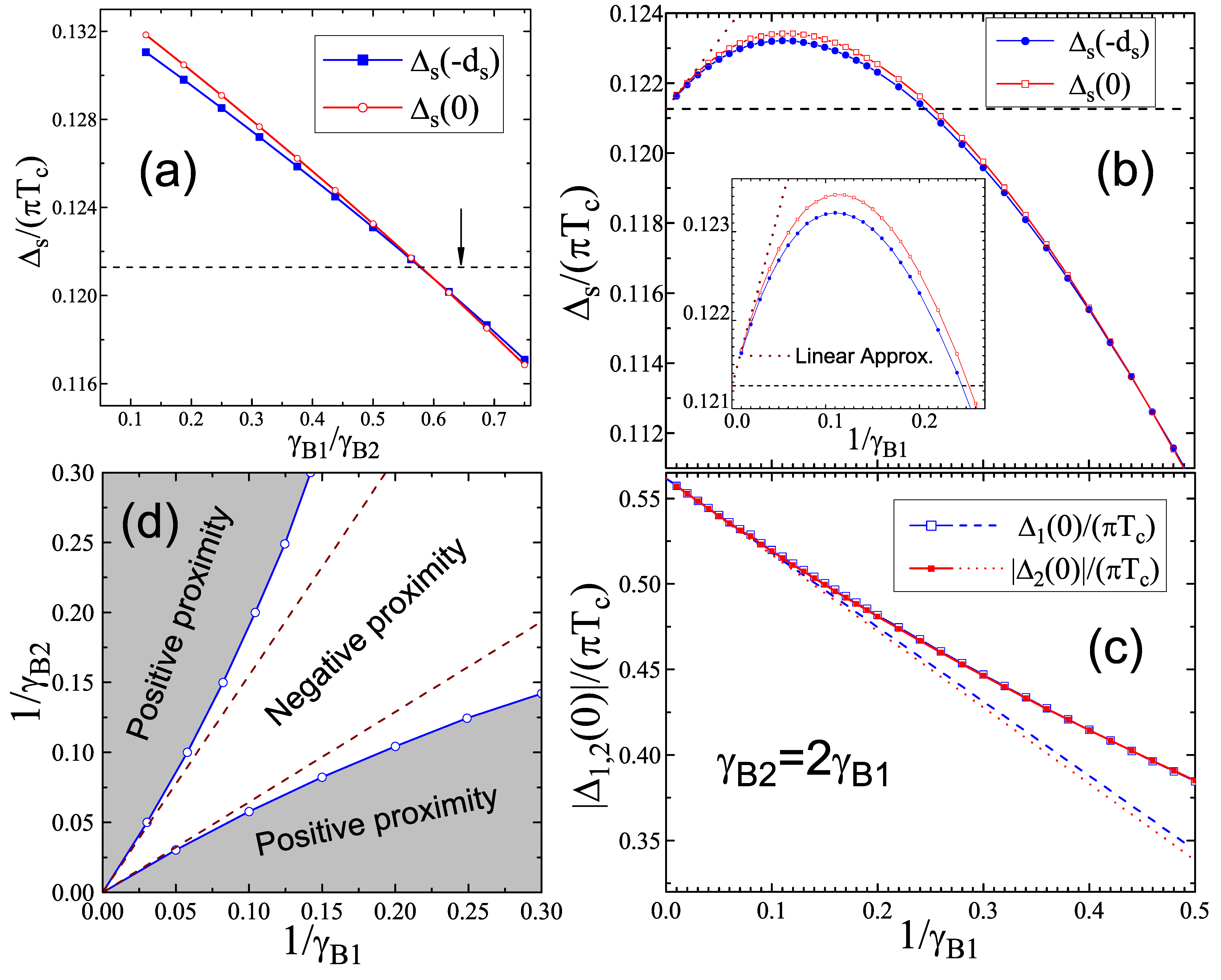}\caption{(Color online) (a)Plot of
$\Delta_{s}(-d_{s})$ and $\Delta_{s}(0))$ as a
function of the ratio $\gamma_{B1}/\gamma_{B2}$. As $\gamma_{B1}/\gamma_{B2}$
approaches $1$, the enhancement on the $s$-wave side is
replaced by suppression (the proximity effect changes from positive to
negative).  The arrow indicate the expected location of the transition
to the negative proximity from the weak-coupling approximation, $\gamma_{B1}/\gamma_{B2}\approx 0.645$.
(b,c) The dependences of
the $s$-wave gap parameter at the boundaries (plot (b)) and the $s_{\pm}$
gaps at the interface (plot (c)) on the coupling strength with the first
band, $1/\gamma_{B1}$ for fixed ratio $\gamma_{B2}/\gamma_{B1}=2$. The inset in the plot (b)
blows up the small-coupling region. Behavior of the gaps expected from the
linear approximation is also shown (the dotted line in the plot (b) and the
dashed and dotted lines in the plot (c)).
(d) The regions of the negative and positive proximity in the $1/\gamma_{B1}$-$1/\gamma_{B2}$
plane obtained by numerical calculations. The dashed lines show boundaries obtained
within weak-coupling approximation.
Parameters used in the calculation for all plots are $\gamma_{B1}=10$, $\gamma_{\alpha}=10$,
$d_{s}=\xi^{\ast}_{s}$.
} \label{NegPosIdent}
\end{figure}

Now we proceed to systematically study the interplay between the different physical
parameters and the transition from negative to positive proximity effects. We plot
on Fig.\ \ref{NegPosIdent}(a) the value of $\Delta _{s}(-d_{s})$ and $\Delta_{s}(0)$
as a function of the ratio $\gamma_{B1}/\gamma_{B2}$ (the model, of course, is
symmetric with respect to the exchange $\gamma_{B1}\!
\leftrightarrow\!\gamma_{B2}$). $\Delta_{s}$ is suppressed as this ratio gets closer
to one (and the frustration increases), in agreement with our previous qualitative
arguments and analytical calculations. With increasing the ratio the proximity
effect turns from positive to negative when the ratio $\gamma_{B1}/\gamma_{B2}$
exceeds the critical value $\approx 0.578$. The linear approximation for the average
correction to the s-wave gap \eqref{AvDeltasLowT} in the case
$|\Delta_{20}|=\Delta_{10}$ gives the following estimate for this critical value,
$\gamma_{B1}/\gamma_{B2}\approx
(\Delta_{10}-\Delta_{s0})/(\Delta_{10}+\Delta_{s0})$. For our parameters this gives
$\gamma_{B1}/\gamma_{B2}\approx 0.645$ which somewhat exceeds the value obtained in
numerical calculations. As the system gets closer to the line of maximal frustration
$\gamma_{B1}\!=\!\gamma_{B2}$ it eventually undergoes a phase transition to the TRSB
state.

On Figs.\ \ref{NegPosIdent}(b) and \ref{NegPosIdent}(c) we illustrate the dependence
of the gap parameters on the coupling strength $1/\gamma_{B1}$ for fixed ratio
$\gamma_{B2}/\gamma_{B1}= 2$. At small coupling strength, $1/\gamma_{B1}\lesssim
0.07$ behavior of all gaps agrees with the linear approximation described in Sec.
\ref{Sec:AnalytWeak}. At larger coupling the $s$-wave gap strongly deviates from the
linear approximation. In particular, it reaches maximum at some value of coupling
and decreases at larger values. With further increase of coupling, $\Delta_s$ drops
below the bulk value, i.e., proximity becomes negative. Qualitatively, this behavior
can be understood as follows: The positive and negative contributions to $\Delta_s$
are roughly proportional to $\Delta_{1}(0)\!-\!\Delta_s$ and
$|\Delta_{2}(0)\!+\!\Delta_s|$ correspondingly. As the $s_{\pm}$ gap parameters
reduce and become closer to the $s$-wave gap with increasing coupling, the negative
term becomes relatively stronger. Another notable property is that the absolute
values of the $s_{\pm}$ gap parameters $\Delta_1$ and $|\Delta_2|$ remain very
close, in spite of significant asymmetry in coupling strength.

On Fig.\ \ref{NegPosIdent}(d) we show the regions of the negative and positive
proximity in the coupling-strengths plane obtained by numerical calculations and
compare them with predictions of the weak-coupling approximation. In agreement with
the plots \ref{NegPosIdent}(a-c), we can see that stronger coupling favors negative
proximity meaning that the weak-coupling approach underestimates the width of the
negative-proximity region for small $\gamma_{B\alpha}$.

\begin{figure}[ptb]
\includegraphics[width=0.4\textwidth]{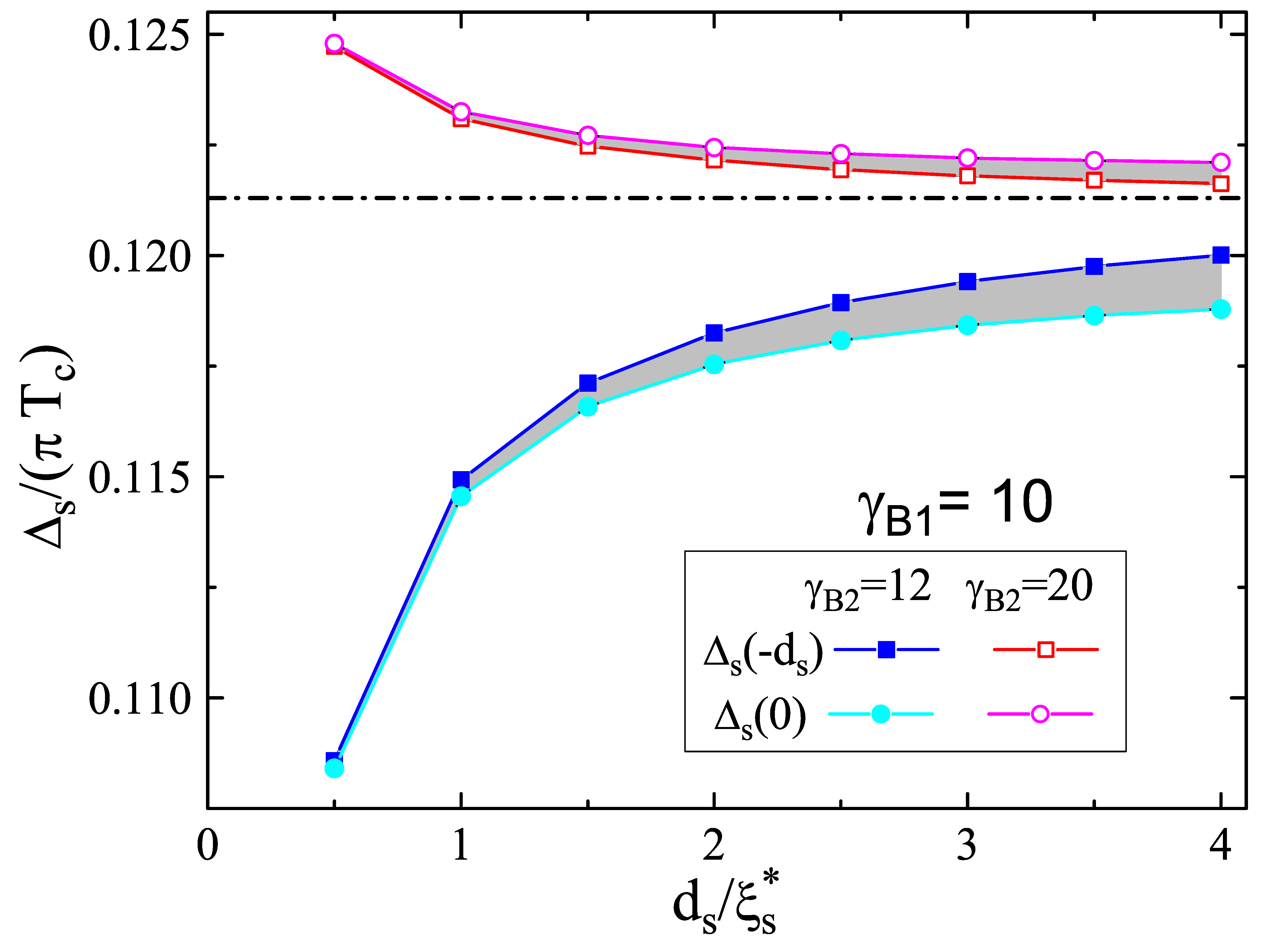}
\caption{(Color online) Plot of $\Delta_{s}(-d_{s})$ and $\Delta_{s}(0)$ as a function of
$d_{s}$ for $\gamma_{B1}=10$ and two values of $\gamma_{B2}$, $12$ and $20$,
corresponding to negative and positive proximity effect.
As the thickness increases $\Delta_{s}$ approaches its bulk value.}
\label{DSdsds}
\end{figure}
Unfortunately, there is no obvious way to control the coupling parameters
$\gamma_{B\alpha}$. One parameter which can be varied relatively easy in experiment
is the thickness of the layers. On Fig.\ \ref{DSdsds} we show the $\Delta_{s}$ as a
function of $d_{s}$ for $\gamma_{B1}=10$ and two values of $\gamma_{B2}$, $12$ and
$20$, corresponding to negative and positive proximity effect. There are two natural
tendencies which can be observed. First, with increase of $d_{s}$ the value of
$\Delta_{s}$ increases (decreases) for the positive (negative) proximity. Second,
the difference between $\Delta_{s}(0)$ and $\Delta_{s}(-d_{s})$ increases with
thickness until it finally saturates for $d_{s}\gg\xi_{s}$.

In this section we demonstrated that both negative and positive proximity effects
may present close to $s$-$s_{\pm}$ interfaces for the aligned state. The sign of the
proximity effect is determined by non-trivial interplay of the physical parameters
in the system. In general, negative proximity effect is expected in the region
around the line which separates the aligned and the TRSB states, for intermediate
frustration of the interface.

\subsection{Nonequal $s_{\pm}$ gaps}
\label{Sec-AlDiffBands}

\begin{figure}[ptb]
\includegraphics[width=0.5\textwidth]{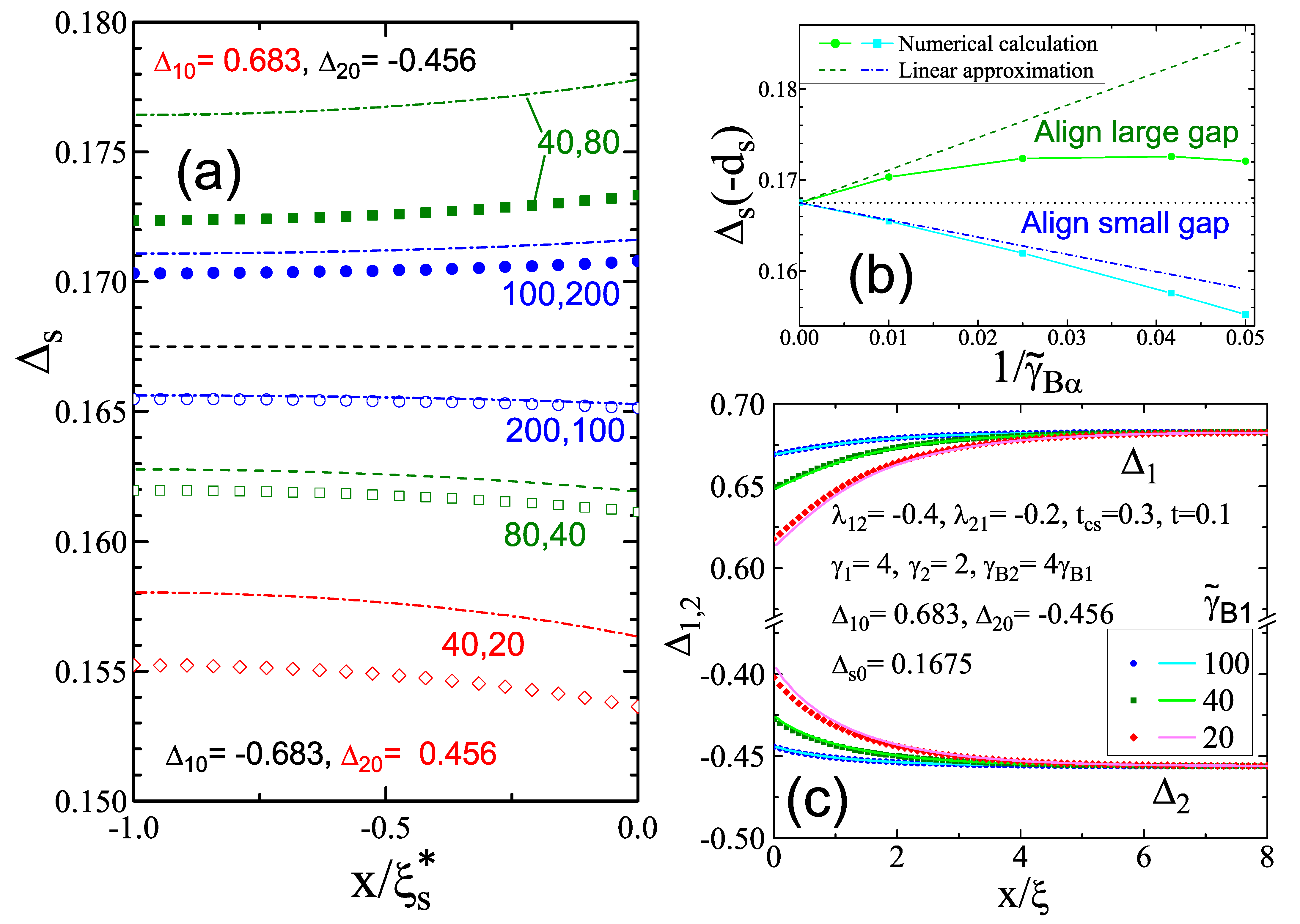}\caption{(Color online) (a)The
coordinate dependences of the $s$-wave order parameter $\Delta_{s}$ for
different couplings to the $s_{\pm}$-bands. The plots are labeled by
$\tilde{\gamma}_{B1},\tilde{\gamma}_{B2}$. Other parameters are listed in the plot (c). The
horizontal dashed line shows bulk gap. The curves above this line (positive
proximity) correspond to the alignment with the large gap, $\Delta_{1}$, while
the curves below this line (negative proximity) correspond to the alignment
with the small gap, $\Delta_{2}$. Three dot-dashed line above and three dashed
lines below show analytical predictions based on linear approximation,
Eq.\ \eqref{Deltasm}. (b) The dependences of the $s$-wave gap at the outside
boundary on the coupling strength with the aligned gap $1/\tilde{\gamma}_{B\alpha}$,
dashed lines show linear approximations. (c) The coordinate dependences of the
$s_{\pm}$-wave order parameters for the alignment with the larger gap. The
solid lines show the linear approximation.}
\label{Fig-DeltaWeakCouplComp}
\end{figure}

In this section we present numerical results illustrating properties of aligned
state when the bulk gaps of $s_{\pm}$ superconductor have different magnitudes. This
situation is probably more typical and has one definite practical advantage with
respect to the case of identical or close $s_{\pm}$ gaps -- different gaps induce
features into the $s$-wave DoS which are well separated in energy and thus easier to
detect experimentally. Therefore, in this section, in addition to behavior of the
gaps, we also study behavior of DoSs. We investigate in detail the shapes of DoS
features and their sensitivity to the coupling parameters.

We first consider the case of weak coupling (large $\gamma_{B\alpha}$ and
$\tilde{\gamma}_{B\alpha}$) and compare numerical calculations with the analytical
results presented of Sec.\ \ref{Sec:AnalytWeak}. In particular, this allows us to
evaluate limits of the weak-coupling approximation. Fig.\
\ref{Fig-DeltaWeakCouplComp} illustrates behavior of the gaps for different coupling
strengths. For $s_{\pm}$ superconductor we consider again the interband coupling
model and the gap values are fixed by the coupling constants which we take as
$\lambda _{12}=-0.4$ and $\lambda_{21}=-0.2$ giving the bulk gaps
$|\Delta_{10}|=0.683$ and $|\Delta_{20}|=0.456$ (in units of $\pi T_{c}$). We assume
$t_{cs} =T_{c}^{s}/T_{c}=0.3$ giving $\Delta_{s0}=0.1675$. We again assume identical
coherence lengths for two $s_{\pm}$-bands. Due to relation \eqref{gamma-a-ratio}, we
have $\gamma_1/\gamma_2=\lambda_{12}/\lambda_{21}=2$ and therefore we select
$\gamma_1=4$, $\gamma_2=2$.  All parameters used in these calculations are listed in
the plot \ref{Fig-DeltaWeakCouplComp}(c). We consider both cases of alignment of the
$s$-wave gap $\Delta_{s}$ with large gap $\Delta_{1}$ ($\Delta_{s}
\!\upuparrows\!\Delta_{1}$) and with small gap $\Delta_{2}$ ($\Delta_{s}
\!\upuparrows\!\Delta_{2}$). We only vary coupling strengths
$\propto1/\tilde{\gamma}_{B\alpha}$ preserving the ratio $\tilde{\gamma}_{B2}
/\tilde{\gamma}_{B1}=2$ for $\Delta_{s} \!\upuparrows\!\Delta_{1}$ and
$\tilde{\gamma} _{B1}/\tilde{\gamma}_{B2}=2$ for $\Delta_{s}
\!\upuparrows\!\Delta_{2}$.

Figure \ref{Fig-DeltaWeakCouplComp}(a) shows the coordinate dependences of the
$s$-wave order parameter. The symbols show numerical results and the lines show
analytical results in the linear order with respect to $1/\tilde{\gamma} _{B\alpha}$
based on Eq.\ \eqref{Deltasm}. The bulk gap is shown by the dashed line. For
selected parameters, the alignment with the large gap corresponds to positive
proximity, while the alignment with the small gap corresponds to negative proximity.
We can see that the linear approximation accurately describes behavior of
$\Delta_{s}$ only for weakest coupling $\tilde{\gamma}_{B(1,2)}=100$. It always
overestimates the gap parameters and breaks down already for rather weak coupling
strength. This is even clearer on Fig.\ \ref{Fig-DeltaWeakCouplComp}(b), where the
gap parameter at the outside surface is plotted versus the coupling strength with
the aligned gap. Deviations are especially large in the case of alignment with the
large gap. The found nonmonotonic dependence of $\Delta_s$ is similar to the case of
identical $s_{\pm}$ gaps shown in Fig.\ \ref{NegPosIdent}(b).

Figure \ref{Fig-DeltaWeakCouplComp}(c) shows the coordinate dependences of the
$s_{\pm}$ gaps in the case $\Delta_{s} \!\upuparrows\!\Delta_{1}$. As expected, both
gaps are suppressed at the interface. This suppression grows with increasing
coupling strength, and disappears away from the boundary for a distance of the order
of the coherence length. Behavior for the case $\Delta_{s}
\!\upuparrows\!\Delta_{2}$ is very similar. It is peculiar that, in contrast to the
$s$-wave gap, the linear approximation accurately describes behavior of the
$s_{\pm}$ gap parameters in the whole studied range of parameters.

\begin{figure}[ptb]
\includegraphics[width=0.5\textwidth]{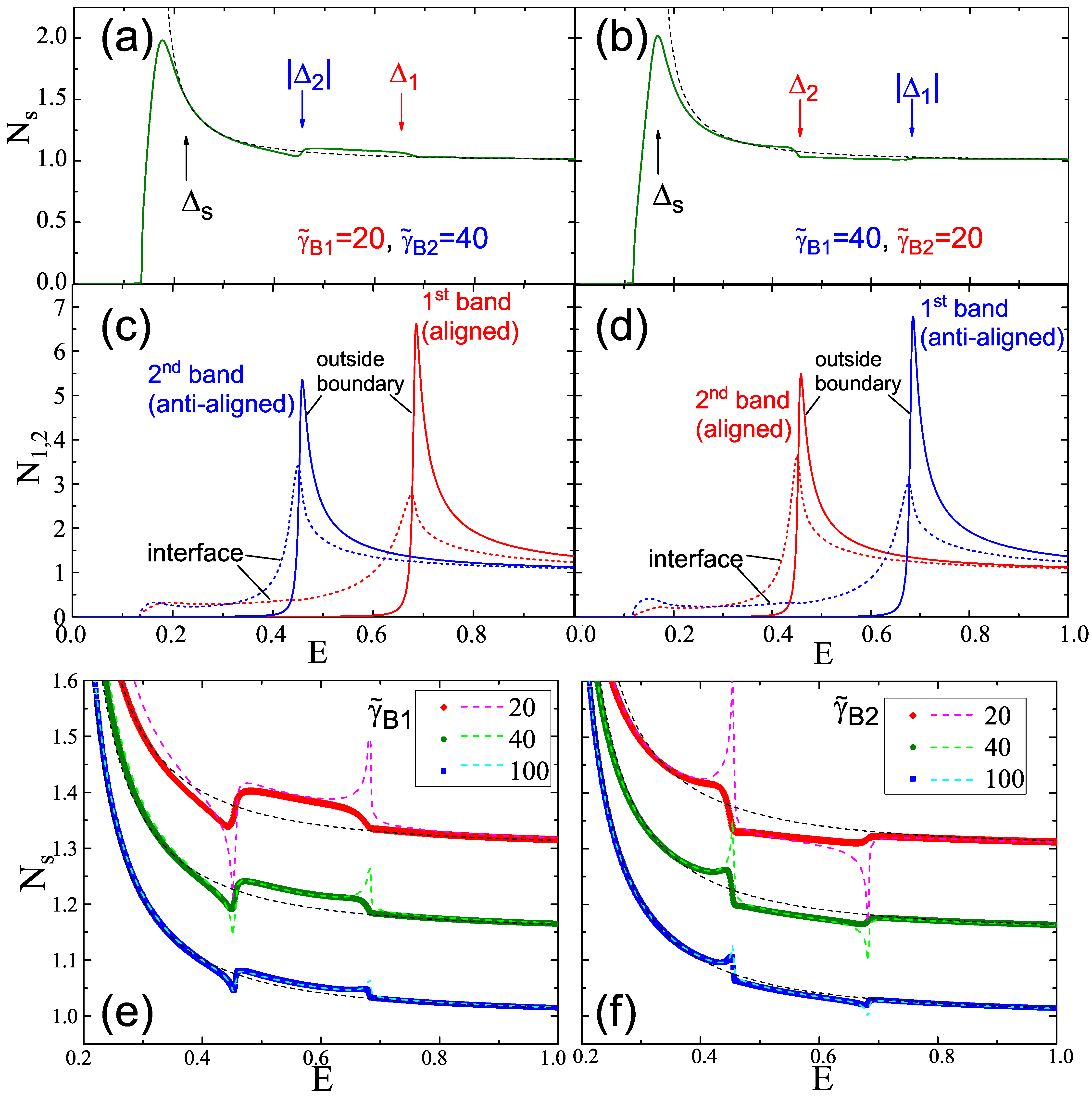}
\caption{(Color online) Behavior of the densities of states for the same parameters as
on Fig.\ \ref{Fig-DeltaWeakCouplComp}. The plots (a) and (b) illustrate
shapes of $s$-wave DoS for the cases when the $s$-wave order parameter
is aligned with large and small gap correspondingly for the strongest
used coupling strength. The dashed lines show the BCS density of states.
The plots (c) and (d) show the partial DoS for two $s_{\pm}$ bands at
the interface and at the outside boundary for these two cases. The plots
(e) and (f) demonstrate the evolution of the feature in the $s$-wave DoS
with varying strength of coupling. The plots are displaced vertically
for clarity. The dashed lines show predictions of the linear
approximation.}
\label{Fig-DoSWeakCouplComp}
\end{figure}
\begin{figure}[ptb]
\includegraphics[width=0.49\textwidth]{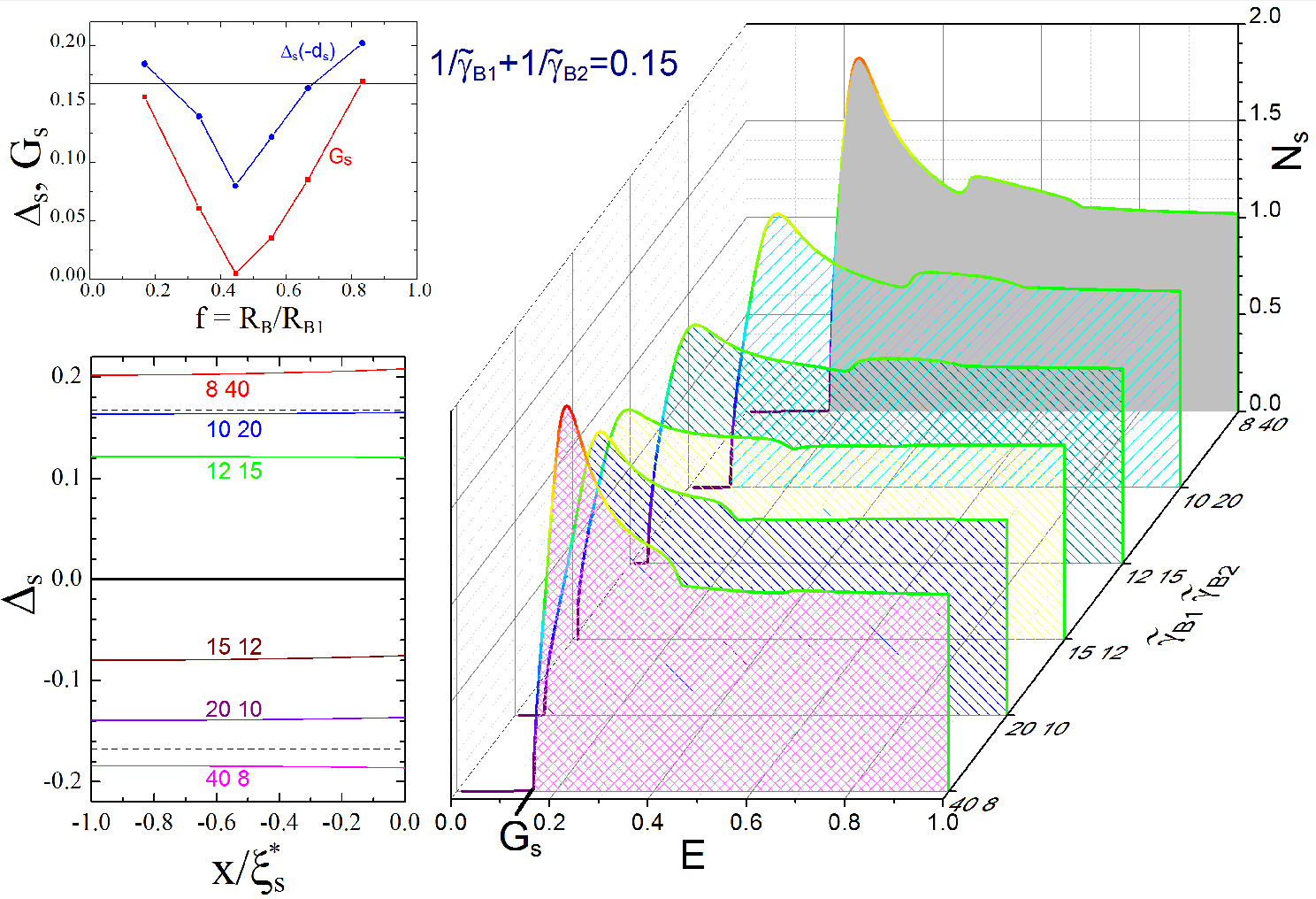}
\caption{(Color online) Evolution of the s-wave the order parameter (left lower plot) and  DoS
(left plot) for the fixed total boundary resistance set by condition
$1/\tilde{\gamma}_{B1}+1/\tilde{\gamma}_{B2}=0.15$ and for varying partial
resistances for two bands. Coupling to the larger gap in the right plot
progressively increases from the back to front plot. Plots are marked by
$\tilde{\gamma}_{B1}$, $\tilde{\gamma}_{B2}$. Other parameters are the same as
in the previous plots. The upper left plot shows the order parameter at the
outside boundary and the gap in spectrum (defined in the right plot) as a
function of a fraction of the first-band conductance with respect to the total
conductance through the boundary.} \label{Fig-NsDs_tzeta0_15}
\end{figure}
Figure \ref{Fig-DoSWeakCouplComp} illustrates behavior of the DoS
 for the same set of parameters. The plots \ref{Fig-DoSWeakCouplComp}(a)
and \ref{Fig-DoSWeakCouplComp}(b) demonstrate full shapes of the $s$-wave DoS at
$x=-d_s$, $N_{s}(E)$, for the cases $\Delta_{s} \!\upuparrows\!\Delta_{1}$ and
$\Delta_{s} \!\upuparrows\!\Delta_{2}$ correspondingly and for strongest coupling in
this series. Interaction with the $s_{\pm}$ superconductor induces specific features
in the $s$-wave DoS near the energies of $s_{\pm}$ gaps. Figures
\ref{Fig-DoSWeakCouplComp}(e,f) show evolution of these $s$-wave DoS features with
increasing coupling strength. The weak-coupling approximation for the case of thin
$s$-superconductor layer, Eq.\ \eqref{sDoScorrWeak}, suggests that the aligned and
anti-aligned gaps should induce a small peak and dip correspondingly
\cite{ProxFingeprintEPL11}. For comparison, the analytical weak-coupling results are
also shown in the plots \ref{Fig-DoSWeakCouplComp}(e,f) by dashed lines. We can see
that, the analytical approximation describes well the shapes of correction except
the regions close to the $s_{\pm}$ gaps where it overestimates the amplitudes of
peaks and dips. Nevertheless, small asymmetric peaks and dips do appear for weak
coupling strength, in agreement with analytical predictions. These peaks and dips
are rapidly smeared with increasing coupling strength. Eq.\ \eqref{sDoScorrWeak}
also suggests that the correction is strongly asymmetric: the $s$-wave DoS is only
enhanced or reduced for $E<|\Delta_{\alpha}|$. This steplike behavior of the
correction is indeed seen in the plots \ref{Fig-DoSWeakCouplComp}(a,b) and, in more
detail, in plots \ref{Fig-DoSWeakCouplComp}(e,f). In fact, with increasing coupling
the peaks and dips evolve into smooth up and down steps. The amplitude of this
steplike feature between $|\Delta_{2}|$ and $|\Delta_{1}| $ monotonically increases
with increasing coupling and its shape is well described by the analytical
approximation.

The plots \ref{Fig-DoSWeakCouplComp}(c) and \ref{Fig-DoSWeakCouplComp}(d) show
partial DoS for two $s_{\pm}$ bands at the interface and at the outside boundary.
Coupling with $s$-wave superconductor leads to considerable smearing of the DoS
peaks at the interface. The most prominent feature is the appearance of the tails
spreading down to the $s$-wave spectral gap. This is a well-known feature which is
always induced in a superconductor by proximity with either metal or weaker
superconductor \cite{McMillanPR68}. Note also that the peaks positions for DoS at
the interface do not shift much with respect the bulk gap values. There is no
qualitative difference in the DoS shapes between the aligned and anti-aligned bands.

We also studied the evolution of the proximity properties with varying the partial
resistances for two $s_{\pm}$ bands, while keeping fixed the total boundary
resistance. As the total boundary \emph{conductance} $\Sigma_{B}=1/R_{B}$ is equal
to sum of the partial boundary conductances $1/R_{B\alpha} \propto
1/\tilde{\gamma}_{B\alpha}$, the total boundary resistance can be set by fixing
$1/\tilde{\gamma}_{B}=1/\tilde{\gamma}_{B1}+1/\tilde{\gamma}_{B2}$. We select this
parameter as $1/\tilde{\gamma}_{B}=0.15$ corresponding to the moderately strong
coupling strength.  All other parameters are the same as in Fig.\
\ref{Fig-DeltaWeakCouplComp}. Figure \ref{Fig-NsDs_tzeta0_15} illustrates evolution
of the s-wave order parameter and DoS for varying partial resistances for two bands.
The plots are marked by the pair $\tilde{\gamma}_{B1},\tilde{\gamma}_{B2}$. We can
see that for strongly asymmetric coupling proximity effect is positive, there is
pronounced BCS peak near the bulk gap, and pronounced steplike features at the
$s_{\pm}$ gaps. The step amplitude is typically larger for smaller gap. As coupling
becomes more symmetric, the proximity turns negative, the BCS peak smears, and the
amplitudes of the proximity-induced features decrease. Also, the gap in the spectrum
corresponding to vanishing of DoS, $G_s$, rapidly decreases as coupling becomes more
symmetric. This is seen more clearly in the upper left plot where the order
parameter and spectral gap are plotted as function of a fraction of the first-band
conductance with respect to the total conductance, $f=R_{B}/R_{B1}$. Both parameters
have quite sharp cusplike dependence on $f$ and the spectral gap practically
vanishes in the minimum. Note that, in contrast to the BCS DoS, the spectral gap is
significantly smaller than the order parameter.

In the next section we consider behavior of the order parameter for the TRSB state.

\section{TRSB state}
\label{TRSB}

For illustration we consider here only the simplest TRSB state, in the case the
two-gap superconductor has identical bands, coupled symmetrically to the $s$-wave
layer (see the panel at the right upper corner of Fig.\ \ref{SchemFig}). The
symmetry simplifies the calculations considerably. This is also the case which
favors the TRSB state most, since the frustration at the interface is the strongest
possible.

Due to the symmetry of the problem we can choose $\Phi_{s}(0)$ to be real. It is
also clear that $\Phi_{1}^{*}\!=\! \Phi_{2}$ and we can write $\Phi_{1}\!=i
\omega\tan\theta e^{i \varphi}$, $\Phi_{2}\!=-i \omega \tan\theta e^{-i \varphi}$.
However, to preserve the uniformity in notation we keep the (now redundant) $\alpha$
indexing -- $\varphi_{1}\!=\!-\varphi_{2}\!=\varphi$,
$\theta_{1}\!=\!-\theta_{2}\!=\theta$. The above implies, of course, that we can
write the gap functions as
 $\Delta_{1}\!=\Delta e^{i \chi}$, $\Delta_{2}\!=-\Delta e^{- i \chi}$.
$\varphi$ and $\chi$ parametrizes the deviation from the $s_{\pm}$ state and we
expect that in the bulk $\varphi\!\rightarrow\! 0$ and $\Phi_{1}$, $\Phi_{2}$ become
purely imaginary. From the boundary conditions it is clear that $\Phi_{s}$ can be
chosen real everywhere (both
$\mathrm{Im}(\Phi_{s}(0))\!=\!\mathrm{Im}(\Phi^{\prime}_{s}(0))\!=\!0$ -- no
imaginary component develops).  In the $\theta$-parametrization the s-wave Green's
function is again determined by Eq.\ \eqref{UeqAlSt}. The equations for the
$s_{\pm}$ side are identical to Eqs.\ \eqref{Ue2}, but there are now only two
independent variables (instead of four). In this case, the boundary conditions
\eqref{BoundCondThetaComplex} can be simplified as
\begin{subequations}
\begin{align*}
&\theta^{\prime}_{\alpha}=\frac{1}{\gamma
   _{B\alpha }}\Big( \sin {\theta _s} \cos {\theta _{\alpha
   }} \sin {\varphi _{\alpha }}+\cos {\theta
   _s} \sin {\theta _{\alpha }}\Big),\nonumber\\
&\varphi^{\prime}_{\alpha}=\frac{1}{\gamma
   _{B\alpha }}\sin {\theta _s} \csc {\theta _{\alpha
   }} \cos {\varphi _{\alpha }},\nonumber\\
&\theta^{\prime}_{s}=-\frac{2}{\gamma_{\alpha}\gamma
   _{B\alpha }} \cos {\theta _s} \sin {\theta _{\alpha
   }} \left(\tan {\theta _s} \cot {\theta
   _{\alpha }}+\sin {\varphi _{\alpha
   }}\right)\nonumber.
\end{align*}
\end{subequations}
\begin{figure}[ptb]
\includegraphics[width=0.5\textwidth]{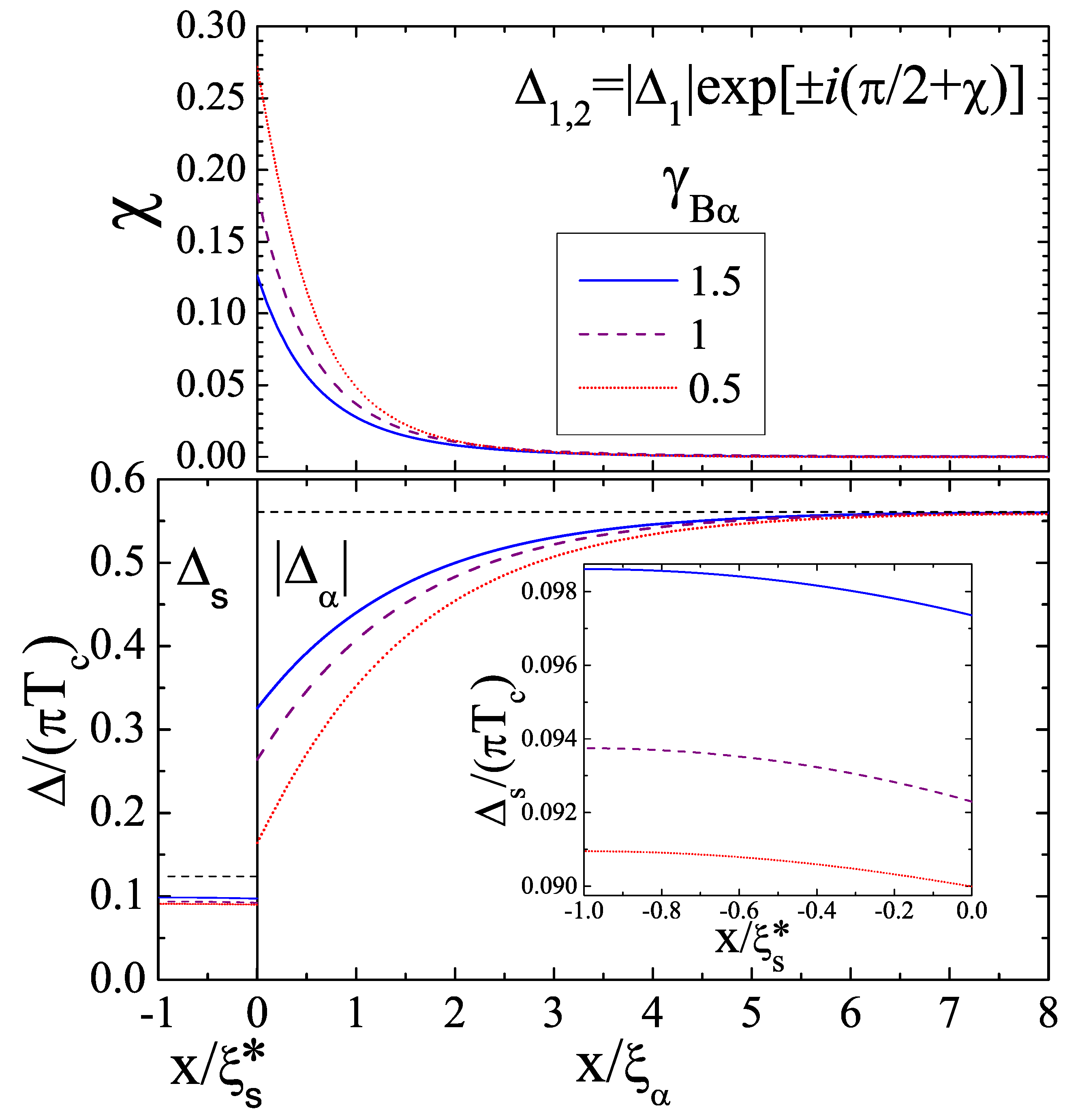}
\caption{(Color online) Behavior of the order parameters in the symmetric TRSB
state for three different boundary transparency parameters
$\gamma_{B\alpha}$ and for $\gamma_{\alpha}=5$. The rest of the
parameters are given in the text. Lower plots show the coordinate
dependences of the absolute values of the order parameters and upper
plot show the phase $\chi(x)$ of the $s_{\pm}$ order parameter.
$\chi$ goes to zero in the bulk, where the state is
$s_{\pm}$, with phase rotated by $\pi/2$ with respect to the
$s$-wave OP. Note that $|\Delta_{1}(0)|$ and $\Delta_{s}$ decrease
and the value of $\chi(0)$ increases with the increase of
the interface transparency.}
\label{DSymCase}
\end{figure}

We proceed to solve the equations numerically. On Fig.\ \ref{DSymCase} we show the
results of calculation for $\Delta$, $\chi$ (we again remind the reader that
$\Delta_{2}=-\Delta_{1}$, $\chi_{2}=-\chi_{1}$) and $\Delta_{s}$ done for different
values of $\gamma_{B \alpha}$. Several things should be noted: First, $\chi(x)$ is
positive and thus the $s_{\pm}$ gaps are pushed away from $\Delta_{s}$, in
accordance with the argument in Section \ref{Sec:Ueq}; Second, several
$\xi_{\alpha}$ away from the boundary,  $\chi$ becomes very small and the order
parameter returns to its $s_{\pm}$ bulk form (since the TRSB state exists because
the frustration created by the boundary); Third, decreasing $\gamma_{B \alpha}$
(increasing the boundary transparency) tends to reduce both $\Delta_{\alpha}(0)$ and
$\Delta_{s}(0)$, and increase $\chi(0)$, as expected. It is important to note that
decreasing $\gamma_{B \alpha}$ pushes $\Delta_{s}(x)$ down {\it everywhere}, while
keeping its general shape intact (unlike the changes of $\Delta_{\alpha}(x)$). This
is due to the two factors, which generally decrease the value of
$\Delta^{\prime}_{\alpha}(0)$, and thus enhance the strength of the proximity effect
away from the interface. The first one is the relative thinness of the $s$-wave
layer, and the other is the large value of $\gamma_{\alpha}$.

\begin{figure}[ptb]
\includegraphics[width=0.5\textwidth]{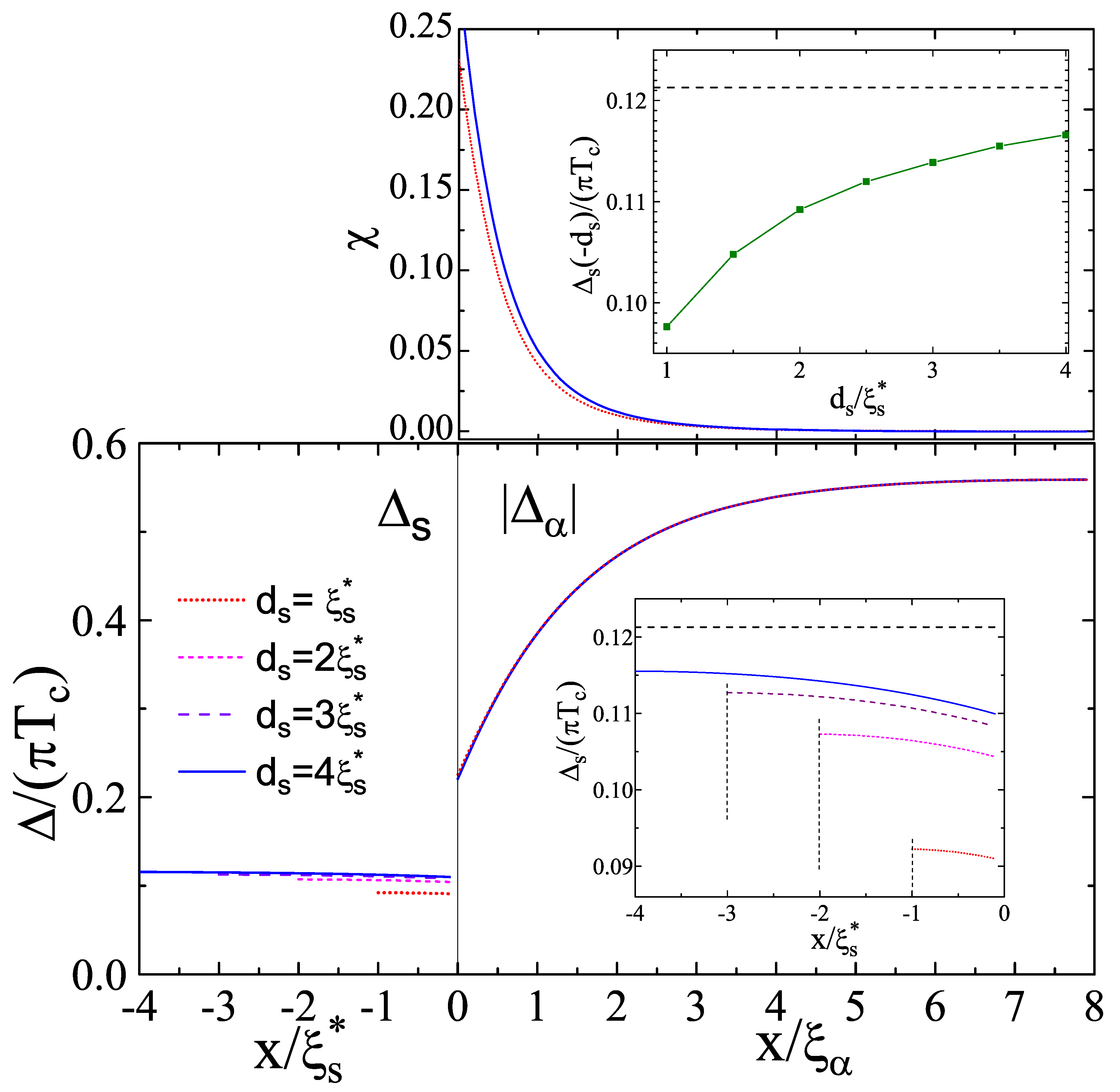}
\caption{(Color online) $\Delta(x)$, $\chi(x)$ and $\Delta_{s}(x)$ for $\gamma_{\alpha}=5,
\gamma_{B\alpha}=0.75\ $ and four different values of $d_s$. The changes on the
right side are generally quite small. In contrast, $\Delta_s(x)$ is suppressed
noticeably for thin films. This is more clearly seen in the lower inset which
shows the blowup plots of $\Delta_{s}(x)$. The upper inset shows the $d_s$
dependence of the $\Delta_{s}$ at the outer surface. } \label{A2symm}
\end{figure}

Since controlling the interface parameters $\gamma_{B \alpha}$ is not easy
experimentally, we also study evolution with increasing thickness of the $s$-wave
superconductor, $d_{s}$, for the same parameters of the interface. We vary the
thickness of the $s$-wave layer from $\xi^{\ast}_s$ to $4\xi^{\ast}_s$ and show the
results of the calculation on Fig.\ \ref{A2symm}. On the $s_{\pm}$ side the changes
are modest -- $\chi$ shifts a bit, while $\Delta_{\alpha}$ is virtually unchanged
for different $d_s$. Note that as $d_s$ decreases, the gap on the right side gets
slightly closer to the bulk $s_{\pm}$ state ($\chi$ goes down). The changes on the
other side are more pronounced. $\Delta_{s}$ is always suppressed as $x\rightarrow
0$, but, as in the previous case, $\Delta_{s}$ is significantly below its bulk value
\emph{everywhere}.

These results are summarized in the upper inset of  Fig.\ \ref{A2symm} where we plot
the value of the gap on the external left boundary as a function of the $s$-wave
layer thickness. We see the result of the negative proximity effect as a suppression
of $\Delta_s(-d_s)$ for thinner films. With the increase of the film thickness
$\Delta_s(-d_s)$ goes up, however, it stays noticeably below its bulk value even for
$d_s=4 \xi^{\ast}_s$. This is again due to the relatively large $\gamma_{\alpha}$,
which enhances the proximity effects away from the interface.

We can see that accurate numerical computations confirm the structure of the TRSB
state expected from general considerations. In particular, in the TRSB state the
negative proximity effect is present and, depending on the precise values of various
physical parameters, can be quite pronounced.

\section{Discussion and Conclusions}
\label{DaC}

One important point should be addressed before we apply our conclusions to
structures involving iron-based superconductors. These materials have quite short
coherence length and are likely in the clean limit. In conventional superconductors
there is smooth interpolation between the clean and the dirty limit and results
obtained by solving the Usadel equations are expected to be qualitatively or even
quantitatively correct even in the clean limit \cite{Golubov3}. Furthermore, the
boundary itself acts as a scatterer, and can push the region up to several $\xi$
away in the dirty regime. This again justifies the use of the Usadel equations. In
$s_{\pm}$ superconductors the situation is much less clear -- because of the
unconventional nature of the order parameter interband impurity scattering is
pair-breaking. If strong, it can completely destabilize the $s_{\pm}$ state
\cite{Efremov}. However, if this scattering channel can be neglected, as we have
done (so $s_{\pm}$ is stable even close to the interface), then our approach should
provide a reasonably good description. Whether this can be justified in realistic
experimental situations is unclear, at least at the moment.

In conclusion, we have studied the proximity effects close to a boundary between $s$
and $s_{\pm}$ superconductors. Based on frustrated Josephson junction model, we have
suggested phase diagram for such system. Because of the frustration, present at such
interface, several interesting phenomena are possible. In the case of maximum
frustration -- when the coupling of the two gaps on the $s_{\pm}$ side with the
$s$-wave superconductor is comparable -- a state which breaks time-reversal symmetry
appears. Such superconducting state is also characterized by a negative proximity
effect - because of the frustration the gap amplitudes on both sides are suppressed.
In the case of significant asymmetry in the inter-boundary coupling $s_{\pm}$ can be
stabilized even close to the interface, but becomes asymmetric there. Interestingly,
such aligned state may also lead to negative proximity effect. It is very important
to note that this effect is unique to the $s$-$s_{\pm}$ structures, and is not
present close to a conventional $s$-$s_{++}$ boundary. Observation of such effect in
structures with iron pnictides/chalcogenides will be a definitive proof that their
order parameter belongs to the unconventional $s_{\pm}$ class.

\begin{acknowledgments}
We acknowledge useful discussions with Thomas Proslier, Alex Levchenko, Laura
Greene, Dale Van Harlingen, and James Eckstein . This work was supported by UChicago
Argonne, LLC, operator of Argonne National Laboratory, a U.S. Department of Energy
Office of Science laboratory, operated under contract No. DE-AC02-06CH11357, and by
the \textquotedblleft Center for Emergent Superconductivity\textquotedblright, an
Energy Frontier Research Center funded by the U.S. Department of Energy, Office of
Science, Office of Basic Energy Sciences under Award Number DE-AC0298CH1088.
\end{acknowledgments}

\appendix

\section{Boundary-induced corrections in the the weak-coupling limit for aligned state}
\label{App-WeakCoupl}

In this appendix we consider weak-coupling case $\gamma_{B\alpha}\gg1$ and derive
corrections to the Green's functions and gap parameter for aligned state in the
linear order with respect to the coupling parameters $1/\gamma_{B\alpha}$.

\subsection{ $s$-wave superconductor}

The first-order corrections to the $s$-wave Green's functions $\tilde{\Phi} _{s}$
and gap $\tilde{\Delta}_{s}$ obey the following equations
\begin{subequations}
\begin{align}
\xi_{s,\omega}^{2}\tilde{\Phi}_{s}^{\prime\prime}-\tilde{\Phi}_{s}  &  =-
\tilde{\Delta}_{s},\label{UeqsLin}\\
\lambda_{s}2\pi T\sum_{\omega>0}\frac{\omega^{2}\tilde{\Phi}_{s}}{\left(
\omega^{2}+\Delta_{s}^{2}\right)  ^{3/2}}  &  =\tilde{\Delta}_{s}
\label{SelfConsEqLins}
\end{align}
with $\xi_{s,\omega}^{2}=D_{s}/(2\sqrt{\omega^{2}+\Delta_{s}^{2}}
)=\xi_{s,\Delta}^{2}(\Delta_{s}/\sqrt{\omega^{2}+\Delta_{s}^{2}})$ and
$\xi_{s,\Delta}^{2}=D_{s}/\left(  2\Delta_{s}\right)  $. In the boundary condition
we can neglect difference between $\Phi$'s and $\Delta$'s and approximate $\Delta$'s
by their bulk values. This gives
\end{subequations}
\begin{equation}
\xi_{s}^{\ast}G_{s}\tilde{\Phi}_{s}^{\prime}=-\sum_{\alpha}\frac{G_{\alpha}
}{\tilde{\gamma}_{B\alpha}}(\Delta_{s}-\Delta_{\alpha}).\label{BoundCondLins}
\end{equation}
at $x=0$ with $\tilde{\gamma}_{B\alpha}\equiv\gamma_{\alpha}\gamma_{B\alpha}$. The
self-consistency condition also can be rewritten as
\begin{equation}
2\pi T\sum_{\omega>0}\left[  \frac{\omega^{2}\tilde{\Phi}_{s}}{\left(
\omega^{2}+\Delta_{s}^{2}\right)  ^{3/2}}-\frac{\tilde{\Delta}_{s}}{\omega}
\right]  +\tilde{\Delta}_{s}\ln\frac{T_{c}^{s}}{T}=0.\label{SelfConsEqLin1}
\end{equation}
We split $\tilde{\Phi}_{s}$ into the two contributions, $\tilde{\Phi} _{s}=
\tilde{\Phi}_{s,b}+\tilde{\Phi}_{s,\Delta}$, where $\tilde{\Phi}_{s,b}$ is induced
by the boundary condition and $\tilde{\Phi}_{s,\Delta}$ is induced by the gap
correction. The first contribution $\tilde{\Phi}_{s,b}$ can be found from the
following equation and the boundary condition
\begin{subequations}
\begin{align}
\xi_{s,\omega}^{2}\tilde{\Phi}_{s,b}^{\prime\prime}-\tilde{\Phi}_{s,b}  &
=0,\label{EqPhi_sb}\\
\xi_{s}^{\ast}G_{s}\Phi_{s,b}^{\prime}  &  =-\sum_{\alpha}\frac{G_{\alpha} }{
\tilde{\gamma}_{B\alpha}}(\Delta_{s}-\Delta_{\alpha}),\label{BCPhi_sb0}
\end{align}
while the second contribution $\tilde{\Phi}_{s,\Delta}$ obeys the following equation
and the boundary condition
\end{subequations}
\begin{subequations}
\begin{align}
\xi_{s,\omega}^{2}\tilde{\Phi}_{s,\Delta}^{\prime\prime}-\tilde{\Phi}
_{s,\Delta}  &  =-\tilde{\Delta}_{s},\label{EqPhiD}\\
\xi_{s}^{\ast}G_{s}\Phi_{s,\Delta}^{\prime}  &  =0.\label{BCPhiD}
\end{align}
The solution $\tilde{\Phi}_{s,b}(x)$ of the linear equation (\ref{EqPhi_sb}) with
the boundary condition $\tilde{\Phi}_{s,b}^{\prime}=0$ at $x=-d_{s}$ is given by
\end{subequations}
\begin{equation}
\tilde{\Phi}_{s,b}(x)=C_{s,b}\cosh\left(  \frac{x+d_{s}}{\xi_{s,\omega} }
\right)  ,\label{Phi-sbSol}
\end{equation}
where the constant $C_{s,b}$ can be found from the boundary condition at $x=0 $
\[
C_{s,b}=-\frac{\xi_{s,\omega}/\xi_{s}^{\ast}}{\sinh\left(  d_{s}/\xi
_{s,\omega}\right)  }\sum_{\alpha}\frac{G_{\alpha}}{\tilde{\gamma}_{B\alpha
}G_{s}}(\Delta_{s}-\Delta_{\alpha}).
\]
We compute $\tilde{\Phi}_{s,\Delta}$ and $\tilde{\Delta}_{s}$ using the Fourier
expansion, $\tilde{\Phi}_{s,\Delta}=\sum_{m=0}^{\infty}\tilde{\Phi}
_{s,\Delta,m}\cos k_{m}x$,$\ \tilde{\Delta}_{s}=\sum_{m=0}^{\infty} \tilde{
\Delta}_{s,m}\cos k_{m}x$ with $k_{m}=m\pi/d_{s}$. For $\tilde{\Phi} _{s,b}(x)$,
Eq.\ (\ref{Phi-sbSol}), the Fourier components $\tilde{\Phi} _{s,b,m}$ can be
computed explicitly
\begin{equation}
\tilde{\Phi}_{s,b,m}=-\frac{(2\!-\!\delta_{m})\xi_{s,\omega}^{2}/(d_{s}\xi
_{s}^{\ast})}{1+\xi_{s,\omega}^{2}k_{m}^{2}}\sum_{\alpha}\frac{\sqrt{
\omega^{2}\!+\!\Delta_{s}^{2}}}{\sqrt{\omega^{2}\!+\!\Delta_{\alpha}^{2}}} \frac{
\Delta_{s}\!-\!\Delta_{\alpha}}{\tilde{\gamma}_{B\alpha}}.\label{Phisbm}
\end{equation}
Eq. (\ref{EqPhiD}) immediately gives the following relation between the Fourier
components $\tilde{\Phi}_{s,\Delta,m}$ and $\tilde{\Delta}_{s,m}$
\begin{equation}
\tilde{\Phi}_{s,\Delta,m}=\frac{\tilde{\Delta}_{s,m}}{1+\xi_{s,\omega}
^{2}k_{m}^{2}}.\label{PhisDm-Dm}
\end{equation}
Substituting this result into the self-consistency condition (\ref{SelfConsEqLins}),
we express $\tilde{\Delta}_{s,m}$ via $\tilde{\Phi} _{s,b,m}$
\begin{align}
\tilde{\Delta}_{s,m}  &  =\left\{  2\pi T\sum_{\omega>0}\frac{\Delta_{s}
^{2}+\omega^{2}\frac{\xi_{s,\omega}^{2}k_{m}^{2}}{1+\xi_{s,\omega}^{2}
k_{m}^{2}}}{\left(  \omega^{2}+\Delta_{s}^{2}\right)  ^{3/2}}\right\}
^{-1}\nonumber\\
&  \times2\pi T\sum_{\omega>0}\frac{\omega^{2}\tilde{\Phi}_{s,b,m}}{\left(
\omega^{2}+\Delta_{s}^{2}\right)  ^{3/2}}.\label{DmPhisbm}
\end{align}
Eqs. (\ref{Phisbm}), (\ref{PhisDm-Dm}), and (\ref{DmPhisbm}) provide a complete
solution of the problem.

For convenient comparison with the numerical calculations, we also present these
results in the reduced form in which $d_{s}$ is measured in units of
$\xi_{s}^{\ast}=D_{s}/(2\pi T_{c})$ and energies ($\omega$, $\Delta$'s and $\Phi$'s)
in units of $\pi T_{c}$ \begin{widetext}
\begin{align}
\tilde{\Delta}_{s,m}  &  =\left\{  2t\sum_{\omega>0}\frac{1}{\left(
\omega^{2}+\Delta_{s}^{2}\right)  ^{3/2}}\left[  \Delta_{s}^{2}+\omega
^{2}\frac{\left(  \pi m/d_{s}\right)  ^{2}}{\sqrt{\omega^{2}+\Delta_{s}^{2}
}+\left(  \pi m/d_{s}\right)  ^{2}}\right]  \right\}  ^{-1}\nonumber\\
&  \times2t\sum_{\alpha,\omega>0}\frac{\omega^{2}}{\left(  \omega^{2}
+\Delta_{s}^{2}\right)  \sqrt{\omega^{2}+\Delta_{\alpha}^{2}}}\frac
{(2-\delta_{m})/d_{s}}{\sqrt{\omega^{2}+\Delta_{s}^{2}}+\left(  \pi
m/d_{s}\right)  ^{2}}\frac{\Delta_{\alpha}-\Delta_{s}}{\tilde{\gamma}
_{B\alpha}},\label{DsCorrReduced}\\
\tilde{\Phi}_{s,m}  &  =\frac{1}{1+\left(  \pi m/d_{s}\right)  ^{2}
/\sqrt{\omega^{2}+\Delta_{s}^{2}}}\left(  \tilde{\Delta}_{s,m}+\frac
{2-\delta_{m}}{d_{s}}\sum_{\alpha}\frac{1}{\sqrt{\omega^{2}+\Delta_{\alpha
}^{2}}}\frac{\Delta_{\alpha}-\Delta_{s}}{\tilde{\gamma}_{B\alpha}}\right)
\label{PhisCorrReduced}
\end{align}
\end{widetext}
with $t=T/T_{c}$ and $\omega=2t(n+1/2)$. At low temperatures the summation with
respect to the Matsubara frequencies can be replaced by the integration $2\pi
T\sum_{\omega>0}\rightarrow\int_{0}^{\infty}d\omega$. In this limit we can obtain an
analytical result for the average correction to the order parameter,
$\tilde{\Delta}_{s,0}$,
\begin{align}
\frac{\tilde{\Delta}_{s,0}}{\pi T_{c}}  &  =\frac{\xi_{s}^{\ast}}{d_{s}}
\sum_{\alpha}U(\Delta_{s}/|\Delta_{\alpha}|)\frac{\Delta_{\alpha}-\Delta_{s}
}{\tilde{\gamma}_{B\alpha}|\Delta_{\alpha}|}\label{AvDeltasLowTApp}\\
\text{with }U(a)  &  =\int_{0}^{\infty}dz\frac{z^{2}}{\left(  z^{2}+1\right)
^{3/2}\sqrt{a^{2}z^{2}+1}}\nonumber\\
&  =\frac{K(1-a^{2})-E(1-a^{2})}{1-a^{2}},\nonumber
\end{align}
where $K(m)=\int_{0}^{\pi/2}(1-m\sin^{2}\theta)^{-1/2}d\theta$ and
$E(m)=\int_{0}^{\pi/2}(1-m\sin^{2}\theta)^{1/2}d\theta$ are the complete elliptic
integrals.

We present also the simple analytical results for important particular case of (i)
thin $s$-layer, $d_{s}\ll\xi_{s,\Delta}$, (ii) weaker $s$ -superconductor,
$\Delta_{s}\ll|\Delta_{\alpha}|$, and (iii) low temperatures, $T\ll T_{c}^{s} $. Due
to the first condition, the dominating contribution to the gap correction is given
by the coordinate independent part $\tilde{\Delta}_{s,0}$, which is determined by
the general formula (\ref{AvDeltasLowTApp}). In the limit of
$\Delta_{s}\ll|\Delta_{\alpha}|$ we can use the asymptotics of the function $U(a)$
in the limit $a\ll1$, $U(a)\approx\ln\left(  4/a\right)  -1$ leading to the
following simple result
\begin{equation}
\frac{\tilde{\Delta}_{s,0}}{\pi T_{c}}\approx\frac{\xi_{s}^{\ast}}{d_{s}}
\sum_{\alpha}\frac{\Delta_{\alpha}-\Delta_{s}}{\tilde{\gamma}_{B\alpha}
|\Delta_{\alpha}|}\left[  \ln\left(  \frac{4|\Delta_{\alpha}|}{\Delta_{s} }
\right)  -1\right].\label{DsLimCaseApp}
\end{equation}
The sign of $\tilde{\Delta}_{s,0}$ determines net effect of the $s_{\pm}$
superconductor on the $s$ superconductor (positive vs negative proximity). As one
can expect, the gaps aligned with $\Delta_{s}$ enhance superconductivity and the
gaps anti-aligned with $\Delta_{s}$ suppress superconductivity in the $s$
superconductor. The relative contributions are mostly determined by the electrical
coupling between $s$-superconductor and the $s_{\pm}$ bands.

Weak spatial dependence of $\tilde{\Delta}_{s}$ is determined by the components
$\tilde{\Delta}_{s,m}$ with $m>0$. At $T=0$ these components can be presented as
\begin{equation}
\tilde{\Delta}_{s,m}=-\frac{1}{W_{m}}\frac{2\xi_{s,\Delta}^{2}}{d_{s}\xi_{s}
}\sum_{\alpha}J_{\alpha,m}\frac{\Delta_{s}-\Delta_{\alpha}}{\tilde{\gamma}
_{B\alpha}}\label{DsxLimCaseGen}
\end{equation}
with
\begin{align*}
W_{m}  &  =\int_{0}^{\infty}\frac{dz}{(z^{2}+1)^{3/2}}\left[  1+\frac{
\beta_{m}^{2}z^{2}}{\sqrt{z^{2}+\Delta_{s}^{2}}+\beta_{m}^{2}}\right]  ,\\
J_{\alpha,m}  &  =\int_{0}^{\infty}dz\frac{z^{2}}{\left(  z^{2}+1\right)
\sqrt{ z^{2}+(\Delta_{\alpha}/\Delta_{s})^{2}}}\frac{1}{\sqrt{z^{2}+1}
+\beta_{m}},
\end{align*}
and $\beta_{m}=\left(  \pi m\xi_{s,\Delta}/d_{s}\right)  ^{2}$. The first integral
can be actually evaluated in general case
\[
W_{m}=\frac{1}{\beta_{m}}\left[  \frac{\pi}{2}+\sqrt{\beta_{m}^{2}-1}
\ln\left(  \sqrt{\beta_{m}^{2}-1}+\beta_{m}\right)  \right]
\]
and it has asymptotics $W_{m}\approx\ln2\beta_{m}$ for $\beta_{m}\gg1$. In the
limits $|\Delta_{\alpha}|\gg\Delta_{s}$ and $\Delta_{s}\beta_{m}\gg
|\Delta_{\alpha}|$ the second integral can be evaluated as
\begin{align*}
J_{\alpha,m}  &  \approx\int_{0}^{\infty}dz\frac{1}{\sqrt{z^{2}+(\Delta
_{\alpha}/\Delta_{s})^{2}}}\frac{1}{z+\beta_{m}}\\
&  \approx\frac{1}{\beta_{m}}\ln\frac{2\Delta_{s}\beta_{m}}{|\Delta_{\alpha}|}
\end{align*}
Collecting terms, we obtain
\begin{equation}
\tilde{\Delta}_{s,m}\approx-\sum_{\alpha}\left[  1+\frac{\ln\left(  \Delta
_{s}/|\Delta_{\alpha}|\right)  }{\ln\left[  2\left(  \pi m\xi_{s,\Delta}
/d_{s}\right)  ^{2}\right]  }\right]  \frac{2d_{s}/\xi_{s}^{\ast}}{\left(  \pi
m\right)  ^{2}}\frac{\Delta_{s}-\Delta_{\alpha}}{\tilde{\gamma}_{B\alpha}
}\label{DsmLim}
\end{equation}
Using relation $(|x|\!-\!1)^{2}\!=\!\frac{1}{3}\!+\!4\sum_{m=1}^{\infty}
\frac{\cos(\pi m x)}{\left(  \pi m\right)  ^{2}}$, we can approximately present the
gap correction in real space as
\begin{align}
\tilde{\Delta}_{s}(x)  &  \approx\tilde{\Delta}_{s,0}-\frac{d_{s}}{\xi
_{s}^{\ast}}\sum_{\alpha}\frac{\Delta_{s}-\Delta_{\alpha}}{\tilde{\gamma}
_{B\alpha}}\left[  \frac{(x+d_{s})^{2}}{2d_{s}^{2}}-\frac{1}{6}\right]
\nonumber\\
&  \times\left[  1+\frac{\ln\left(  \Delta_{s}/|\Delta_{\alpha}|\right)  }{
2\ln\left(  \pi\sqrt{2}\xi_{s,\Delta}/d_{s}\right)  }\right]
\label{DeltasRealLim}
\end{align}
Correspondingly, for the Green's function for the same limits we derive
\begin{align}
\tilde{\Phi}_{s}(\omega, x)  &  \approx\tilde{\Delta}_{s,0}-\pi T_{c}\frac{
\xi_{s}^{\ast}}{d_{s}}\sum_{\alpha}\frac{1}{\sqrt{\omega^{2}+\Delta_{ \alpha
}^{2} }}\frac{\Delta_{s}-\Delta_{\alpha}}{\tilde{\gamma}_{B\alpha}}\nonumber\\
&  \times\left[  1+\frac{1}{2}\left(  \frac{x+d_{s}}{\xi_{s,\omega}}\right)
^{2} \right] \label{PhisLim}
\end{align}

Typically, the spatial dependence of the order parameter is correlated with the sign
of proximity: $\tilde{\Delta}_{s}(x)$ increases towards the boundary for positive
proximity and vice versa. Analyzing Eq. (\ref{DeltasRealLim}), however, we can
conclude that this is not always the case. Indeed, the quadratic term changes sign
roughly when $\left(  \Delta_{1}-\Delta _{s}\right)  /\tilde{\gamma}_{B1}-\left(
\Delta_{s}+|\Delta_{2}|\right) \tilde{\gamma}_{B2}=0$. At this point $\tilde{
\Delta}_{s,0}\propto\ln\left( C\Delta_{1}/\Delta_{s}\right)  /\Delta_{1}-\ln\left(
C|\Delta_{2}|/\Delta _{s}\right)  /|\Delta_{2}|$ and it is negative$\ $for$\
\Delta_{1}>|\Delta _{2}|$ meaning that in some narrow range of parameters
$\tilde{\Delta}_{s}(x)$ will increase towards the boundary in spite of negative
proximity effect.

\subsection{$s_{\pm}$-wave superconductor}

We can evaluate corrections to the $s_{\pm}$ gap parameter and Green's functions
following the same general route. The first-order correction to $\Phi_{\alpha}$ with
respect to the coupling strength $\gamma_{B\alpha}^{-1}$ is determined by the
following equation and boundary conditions,
\begin{align}
\frac{D_{\alpha}}{2\omega}G_{\alpha}\tilde{\Phi}_{\alpha}^{\prime\prime}-
\tilde{\Phi}_{\alpha}  &  =-\tilde{\Delta}_{\alpha},\label{UeqPMLin}\\
\xi_{\alpha}G_{\alpha}\Phi_{\alpha}^{\prime}=\frac{G_{s}}{\gamma_{B\alpha} }
(\Delta_{s}-\Delta_{\alpha})\text{, }  &  \text{at }x=0.\label{BoundConPMLin}
\end{align}
with $G_{\alpha}\approx\omega/\sqrt{\omega^{2}+\Delta_{\alpha}^{2}}$ and
$\Phi_{\alpha}^{\prime}=0$ at $x=-d_{\mathrm{\pm}}$. The self-consistency condition
for the linear correction
\begin{equation}
2\pi T\sum_{\beta,\ \omega>0}\lambda_{\alpha\beta}\frac{G_{\beta}\tilde{\Phi
}_{\beta}}{\omega}=\tilde{\Delta}_{\alpha}\label{SelfConPMLin}
\end{equation}
can be rewritten as
\begin{align}
2\pi T  &  \sum_{\omega>0}\left[  \frac{\tilde{\Phi}_{\beta}}{\sqrt{\omega
^{2}+\Phi_{\beta}^{2}}}-\frac{\tilde{\Delta}_{\beta}}{\omega}\right]
+\ln\frac{1}{t}\tilde{\Delta}_{\beta}\nonumber\\
&  =\sum_{\alpha}\left(  \lambda_{\beta\alpha}^{-1}-\lambda^{-1}\delta
_{\beta\alpha}\right)  \tilde{\Delta}_{\alpha},\label{SelfConPMLin1}
\end{align}
where $\lambda$ is the largest eigenvalue of the matrix $\lambda_{\alpha\beta }$.

Similar to the $s$-wave case, we can split $\tilde{\Phi}_{\alpha}$ into the
contributions induced by the boundary condition and by the correction to the gap
parameter, $\tilde{\Phi}_{\alpha}=\tilde{\Phi}_{\alpha,b}+\tilde{\Phi}
_{\alpha,\Delta}$. The equation and the boundary condition for $\tilde{\Phi
}_{\alpha,b}$ are
\begin{align}
\xi_{\alpha,\omega}^{2}\tilde{\Phi}_{\alpha,b}^{\prime\prime}-\tilde{ \Phi
}_{\alpha,b} &  =0,\label{Eq-Phi-PMb}\\
\xi_{\alpha}G_{\alpha}\tilde{\Phi}_{\alpha,b}^{\prime} &  =-\frac{G_{s}}{
\gamma_{B\alpha}}(\Delta_{s}-\Delta_{\alpha})\label{BC-PhiPMb}
\end{align}
with $\xi_{\alpha,\omega}^{2}\equiv D_{\alpha}/(2\sqrt{\omega^{2}
+\Delta_{\alpha}^{2}})=\xi_{\alpha,\Delta}^{2}|\Delta_{\alpha}|/ \sqrt
{\omega^{2}+\Delta_{\alpha}^{2}}$ and $\xi_{\alpha,\Delta}^{2}\equiv
D_{\alpha}/(2|\Delta_{\alpha}|)$. The solution for $\tilde{ \Phi}_{\alpha ,b}(x)$ is
given by
\begin{equation}
\tilde{\Phi}_{\alpha,b}(x)=\frac{\xi_{\alpha,\omega}}{\xi_{\alpha}}
\frac{G_{s}}{\gamma_{B\alpha}G_{\alpha}}(\Delta_{s}-\Delta_{\alpha})
\frac{\cosh\left[  (x-d_{\mathrm{\pm}})/\xi_{\alpha,\omega}\right]  }{
\sinh\left(  d_{\pm}/\xi_{s,\omega}\right)  }.\label{Phi-PMbSol}
\end{equation}
The component $\tilde{\Phi}_{\alpha,\Delta}$ has to be found from the following
equations
\begin{align}
\xi_{\alpha,\omega}^{2}\tilde{\Phi}_{\alpha,\Delta}^{\prime\prime}-
\tilde{\Phi}_{\alpha,\Delta} &  =-\tilde{\Delta}_{\alpha},\label{Eq-Phi-PMD}\\
\xi_{\alpha}G_{\alpha}\Phi_{\alpha,\Delta}^{\prime} &  =0.\label{BC-PhiPMD}
\end{align}
We can again find $\tilde{\Phi}_{\alpha,\Delta}(x)$ and $\tilde{\Delta}
_{\alpha}(x)$ using Fourier transform
\[
\tilde{\Phi}_{\alpha,\Delta}=\sum_{m}\tilde{\Phi}_{\alpha,\Delta,m}\cos
q_{m}x,\ \tilde{\Delta}_{\alpha}=\sum_{m}\tilde{\Delta}_{\alpha,m}\cos q_{m}x
\]
with $q_{m}=m\pi/d_{\mathrm{\pm}}$. From Eq. (\ref{Eq-Phi-PMD}) we immediately find
\begin{equation}
\tilde{\Phi}_{\alpha,\Delta,m}=\frac{\tilde{\Delta}_{\alpha,m}}{1+\xi
_{\alpha,\omega}^{2}q_{m}^{2}}.\label{PhiPMDm-Dm}
\end{equation}
Substituting $\tilde{\Phi}_{\alpha,\Delta}$ into the gap equation, we obtain
relation connecting $\tilde{\Delta}_{\alpha,m}$ with $\tilde{\Phi} _{\beta,b,m}$,
Fourier components of $\tilde{\Phi}_{\beta,b}(x)$,
\begin{align}
\tilde{\Delta}_{\alpha,m} &  =2\pi T\sum_{\beta,\omega>0}U_{m,\alpha\beta}
\frac{\omega^{2}\tilde{\Phi}_{\beta,b,m}}{\left(  \omega^{2}+\Delta_{\beta
}^{2}\right)  ^{3/2}},\label{DmPhiPMbm}\\
U_{m,\alpha\beta} &  =\left[  \lambda_{\alpha\beta}^{-1}-\left(  \lambda
^{-1}+\Sigma_{m,\alpha}\right)  \delta_{\alpha\beta}\right]^{-1},\nonumber\\
\Sigma_{m,\alpha} &  =2\pi T\sum_{\omega>0}\left[  \frac{\omega^{2}}{\left(
\omega^{2}\!+\!\Delta_{\alpha}^{2}\right)  ^{3/2}}\frac{1}{1\!+\!\xi
_{\alpha,\omega}^{2}q_{m}^{2}}-\frac{1}{\omega}\right]  +\ln\frac{1}
{t}.\nonumber
\end{align}
Fourier transformation of the result in Eq.(\ref{Phi-PMbSol}) gives
\begin{equation}
\tilde{\Phi}_{s,b,m}=\frac{\left(  2-\delta_{m}\right)  \xi_{\alpha,\omega
}^{2}/\left(  d_{\mathrm{\pm}}\xi_{\alpha}\right)  }{1+\xi_{\alpha,\omega}
^{2}q_{m}^{2}}\frac{\sqrt{\omega^{2}+\Delta_{\alpha}^{2}}}{\sqrt{\omega
^{2}+\Delta_{s}^{2}}}\frac{\Delta_{s}-\Delta_{\alpha}}{\gamma_{B\alpha}
}.\label{PhiPMbm}
\end{equation}
It is convenient to introduce the degenerate matrix, $w_{\alpha\beta}
=\lambda_{\alpha\beta}^{-1}-\lambda^{-1}\delta_{\alpha\beta}$, $w_{11}
w_{22}-w_{12}w_{21}=0$. Explicitly, the elements of this matrix are given by
\begin{align*}
w_{11} &  =\frac{\sqrt{\lambda_{-}^{2}/4+\lambda_{12}\lambda_{21}}-\lambda
_{-}/2}{\det\lambda},\ w_{12}=-\frac{\lambda_{12}}{\det\lambda}\\
w_{22} &  =\frac{\sqrt{\lambda_{-}^{2}/4+\lambda_{12}\lambda_{21}}+\lambda
_{-}/2}{\det\lambda},\
\end{align*}
with $\lambda_{-}\equiv\lambda_{11}-\lambda_{22}$ and $\det\lambda
\equiv\lambda_{11}\lambda_{22}-\lambda_{12}\lambda_{21}$. The matrix
$U_{m,\alpha\beta}=\left[  w_{\alpha\beta}- \Sigma_{m,\alpha}
\delta_{\alpha\beta}\right]  ^{-1}$ can be presented as
\begin{align}
\hat{U}_{m} &  =\frac{1}{D_{U}}
\begin{bmatrix}
w_{22}\!-\!\Sigma_{m,2} & -w_{12}\\
-w_{21} & w_{11}\!-\!\Sigma_{m,1}
\end{bmatrix}
,\label{Uab}\\
D_{U} &  =-\Sigma_{m,2}w_{11}\!-\!\Sigma_{m,1}w_{22}+\Sigma_{m,1}\Sigma
_{m,2}.\nonumber
\end{align}
The equations (\ref{PhiPMDm-Dm}), (\ref{DmPhiPMbm}), and (\ref{PhiPMbm}) provide
full formal solution of the problem.

For convenient comparison with numerical calculations, we present the above results
in the reduced form
\begin{align*}
\tilde{\Phi}_{\alpha,b,m}  &  =\frac{\left(  2-\delta_{m}\right)  (\xi_{\alpha}/d_{
\mathrm{\pm}})\sqrt{\omega^{2}+\Delta_{\alpha}^{2}}}{\left[  \sqrt{\omega
^{2}+\Delta_{\alpha}^{2}}+\left(\frac{m\pi\xi_{\alpha}}{d_{\mathrm{\pm}}}\right)^{2}\right]
\sqrt{\omega^{2}+\Delta_{s}^{2}}}\frac{\Delta_{s}-\Delta_{\alpha}}{
\gamma_{B\alpha}},\\
\tilde{\Phi}_{\alpha,\Delta,m}  &  =\frac{\sqrt{\omega^{2}+\Delta_{\alpha}
^{2} }}{\sqrt{\omega^{2}+\Delta_{\alpha}^{2}}+(m\pi\xi_{\alpha}/d_{\mathrm{
\pm} })^{2} }\tilde{\Delta}_{\alpha,m},\\
\tilde{\Delta}_{\alpha,m}  &  =2t\sum_{\beta,\omega>0}U_{m,\alpha\beta} \frac{
\omega^{2}}{\left(  \omega^{2}+\Delta_{\beta}^{2}\right)  \sqrt{ \omega
^{2}+\Delta_{s}^{2}}}\\
&  \times\frac{2\xi_{\beta}/d_{\mathrm{\pm}}}{\sqrt{\omega^{2}+\Delta_{ \beta
}^{2} }+(m\pi\xi_{\beta}/d_{\mathrm{\pm}})^{2}}\frac{ \Delta_{s}
-\Delta_{\beta}} {\gamma_{B\beta}}.
\end{align*}

\section{Boundary conditions in $\theta$-parametrization}
\label{App:BoundCondTheta}

In this appendix we derive the boundary conditions (BCs) in
$\theta$-parametrization, appropriate for the different SC states we consider. We
start with the multiband generalization of the Kupriyanov-Lukichev BCs:
\begin{subequations}
\begin{align}
&  \xi^{\ast}_{s}G_{s}^{2}\Phi_{s}^{\prime} =\sum_{\alpha}\frac{\xi_{\alpha} }
{ \gamma_{\alpha}}G_{\alpha}^{2}\Phi_{\alpha}^{\prime}\\
&  \xi_{\alpha} G_{\alpha}\Phi_{\alpha}^{\prime} = -\frac{1}{\gamma_{B \alpha}
} G_{s}(\Phi_{s}-\Phi_{\alpha}).
\end{align}
\label{BCs-App}
\end{subequations}
In the aligned case all $\Phi$'s can be chosen real. Thus we have:
\begin{equation}
\Phi_{\alpha}=\omega\tan{\theta_{\alpha}},\ \ \Phi_{s}=\omega\tan{\ \theta_{s}
},\ \ G_{\alpha}=\cos{\theta_{\alpha}},\ \ G_{s}=\cos{\theta_{s}},\nonumber
\end{equation}
where $\theta$'s are also real. With those the BC at the interface for the aligned
state simplify to:
\begin{align}
&  \xi_s^{\ast}\theta^{\prime}_{s}=\sum_{\alpha}\frac{\theta^{\prime}_{\alpha}}{
\gamma_{\alpha}}\\
&  \xi_\alpha\theta^{\prime}_{\alpha}=\frac{1}{\gamma_{B \alpha}}\sin(\theta_{\alpha}-
\theta_{s}).
\end{align}
This corresponds to equations in the reduced units presented in the text.

For the TRSB state the functions $\Phi_{\{s, \alpha\}}$ are complex and we select
the following parametrization $\Phi_{s}=\omega\tan\theta _{s}e^{i\varphi_{s}}$ and
$\Phi_{\alpha}=i\omega\tan\theta_{\alpha} e^{i\varphi_{\alpha}}$. Taking derivatives
\begin{align*}
\Phi_{s}^{\prime}  & =\omega\left(  \frac{\theta_{s}^{\prime}}{\cos^{2}
\theta_{s}}+i\varphi_{s}^{\prime}\tan\theta_{s}\right)  e^{i\varphi_{s}}\\
\Phi_{\alpha}^{\prime}  & =i\omega\left(  \frac{\theta_{\alpha}^{\prime}}
{\cos^{2}\theta_{\alpha}}+i\varphi_{\alpha}^{\prime}\tan\theta_{\alpha
}\right)  e^{i\varphi_{\alpha}},
\end{align*}
substituting them in Eqs.\ \eqref{BCs-App} and separating real and imaginary parts,
we obtain the boundary conditions for general complex case
\begin{subequations}
\begin{align}
&\xi _{\alpha }\theta _{\alpha }^{\prime }\!=\!-\frac{1}{\gamma _{B\alpha }}
\left[ \cos \theta _{\alpha }\sin \theta _{s}\sin \left( \varphi
_{s}\!-\!\varphi _{\alpha }\right)\! -\!\cos {\theta _{s}}\sin \theta _{\alpha }
\right],  \\
&\xi _{\alpha }\varphi _{\alpha }^{\prime }\sin \theta _{\alpha } =\frac{1}{
\gamma _{B\alpha }}\sin \theta _{s}\cos \left( \varphi _{s}-\varphi _{\alpha
}\right),\\
&\xi _{s}^{\ast }\theta _{s}^{\prime }\! =\!-\!\sum_{\alpha }\frac{1}{\tilde{
\gamma}_{B\alpha }}\left[ \cos {\theta _{\alpha }}\sin \theta _{s}\!+\!\cos
\theta _{s}\sin \theta _{\alpha }\sin \left( \varphi _{\alpha }\!-\!\varphi
_{s}\right) \right],  \\
&\xi _{s}^{\ast }\varphi _{s}^{\prime }\sin \theta _{s} =\sum_{\alpha }
\frac{1}{\tilde{\gamma}_{B\alpha }}\sin \theta _{\alpha }\cos \left( \varphi
_{\alpha }-\varphi _{s}\right).
\end{align}
\label{BoundCondThetaComplex-App}
\end{subequations}


\begin{thebibliography}{99}
\bibitem{PhaseSenseNote}The necessary requirement for such experiment is the
existence of two crystal faces from which tunneling is dominated by the electron and
hole bands. This requirement is very hard to meet.
\bibitem{Note-gammaBa}Note that our definitions of $\gamma_{B \alpha}$ is
different from the one used in Ref. \onlinecite{Golubov1}.
\end{thebibliography}
\end{document}